\documentclass[11pt,a4paper]{article}
\usepackage[T1]{fontenc}
\usepackage{lmodern}
\usepackage[margin=1in]{geometry}
\usepackage{amsmath,amssymb}
\usepackage{graphicx}
\usepackage{microtype}
\usepackage{authblk}
\usepackage[hidelinks]{hyperref}
\hypersetup{pdftitle={Noether Symmetries in Scalar--Vector--Two-Form Cosmology with a Topological BF Interaction},pdfauthor={Amin Rezaei Akbarieh and Yusuf Kucukakca},pdfsubject={Scalar field and isotropic BF-coupled Abelian form triplets},pdfkeywords={Noether symmetry; Dark energy; Two-form fields; Topological BF interaction; Exact cosmological solutions}}
\numberwithin{equation}{section}
\newcommand{\Mpl}{M_{\mathrm{Pl}}}
\newcommand{\Hc}{H_{\mathrm{c}}}
\newcommand{\dd}{\mathrm{d}}

\title{Noether Symmetries in Scalar--Vector--Two-Form Cosmology with a Topological BF Interaction}
\author[1]{Amin Rezaei Akbarieh\thanks{Corresponding author. E-mail: \href{mailto:amin.rezaeiakbarieh@kocaeli.edu.tr}{amin.rezaeiakbarieh@kocaeli.edu.tr}. ORCID: \href{https://orcid.org/0000-0002-4638-3751}{0000-0002-4638-3751}.}}
\author[2]{Yusuf Kucukakca\thanks{E-mail: \href{mailto:ykucukakca@akdeniz.edu.tr}{ykucukakca@akdeniz.edu.tr}. ORCID: \href{https://orcid.org/0000-0002-2921-5414}{0000-0002-2921-5414}.}}
\affil[1]{Department of Physics, Kocaeli University, 41001 Izmit, T\"urkiye}
\affil[2]{Department of Physics, Faculty of Science, Akdeniz University, Antalya 07058, Turkey}
\date{}
\begin{document}
\maketitle

\begin{abstract}
We study a canonical scalar field coupled to isotropic triplets of Abelian one-forms and two-forms through scalar-dependent kinetic functions and a constant topological $BF$ interaction. By retaining the lapse in the homogeneous reduction, we recover the Friedmann constraint and identify a sufficient domain in which all twelve physical modes have positive kinetic terms and luminal high-frequency propagation. A separate exponential model admits an accelerating mixed solution with $a\propto t^2$, nonzero asymptotic form fractions, and bounded normalized linear perturbations. For the Noether analysis, we choose a quadratic effective potential in flat minisuperspace coordinates and solve the zero-boundary determining equations. The canonical scalar potential is hyperbolic-squared or exponential, rather than a power law in the scalar field. The causal character of the potential coordinate separates three branches: Branch I admits a de Sitter limit, Branch II cannot sustain acceleration on regular future-complete expanding trajectories, and Branch III gives an exact solution with both form sectors active and late expansion $a\propto t^{4/3}$. Comparisons with consistent uncoupled and reduced-field limits separate the form contributions from the scalar-driven acceleration. The $BF$ term changes the finite-time exchange between the form sectors but supplies no direct negative-pressure source.
\end{abstract}

\noindent\textbf{Keywords:} Noether symmetry; Dark energy; Two-form fields; Topological BF interaction; Exact cosmological solutions.

\section{Introduction}
\label{sec:introduction}

Recent Type Ia supernova compilations and baryon acoustic oscillation measurements constrain the late-time expansion with increasing precision~\cite{Brout:2022vxf,DES:2024jxu,DESI:2025zgx}. A cosmological constant remains a simple reference model, while a rolling scalar field allows the dark-energy density to evolve through its kinetic and potential energies. Exponential potentials provide a useful setting for relating the potential slope to scaling solutions and accelerated expansion~\cite{Copeland1998,Andriot:2024jsh}. Coupling the scalar to additional fields then raises a related question: how do their energy exchange and dynamical constraints affect these solutions?

Vector fields extend this setting, although a nonzero spatial vector generally selects a preferred direction. Isotropy can be preserved by three identical vectors with equal amplitudes along mutually orthogonal directions, a construction that continues to support studies of background evolution and perturbations~\cite{Yamamoto:2012sq,Murata:2021vnb,Chen:2022ccf}. Recent dark-energy models also explore scalar--gauge interactions, gauged quintessence, and vector--fluid exchange~\cite{Gomez:2021jbo,Kaneta:2022kjj,Pookkillath:2024ycd}. Their different coupling prescriptions make it important to specify both the field content and the constraints when comparing their dynamics.

Two-form fields offer a further possibility, with isotropic triplets and their perturbative consistency receiving renewed attention~\cite{Ajith:2022wia,Horii:2025jen}. In four dimensions, a constant-coefficient interaction $B\wedge F$ preserves the Abelian gauge symmetries up to a boundary term and realizes the established mechanism of topological mass generation~\cite{Cremmer:1973mg,Allen:1990gb}. Its place in coupled cosmological form theories and its massive-vector dual are already known~\cite{BeltranAlmeida2019,BeltranAlmeida2020}. Although this interaction has no direct metric stress tensor, it changes the form equations and thereby the evolving energy distribution.

An accelerating background must also be tested at the perturbative level. Recent analyses of nonminimally coupled vector and Kalb--Ramond fields show how longitudinal or other propagating modes can invalidate an apparently acceptable background~\cite{Capanelli:2024pzd,Hell:2026mle}. A related lesson arises in Noether-symmetric Yang--Mills cosmology with a Pontryagin coupling~\cite{Akbarieh2026Pontryagin}. Constraints must therefore be identified before testing the reduced kinetic matrices. Positive kinetic energy, healthy short-wavelength propagation, and bounded perturbations over cosmological times remain distinct requirements in a time-dependent coupled system.

Noether symmetries organize the background equations through conserved quantities and coordinates adapted to the dynamics. Recent work develops this approach in multifield and scalar--tensor cosmology~\cite{Giacomini2022Multifield,Miranda2024Horndeski}, while applications to vector dark energy and teleparallel vector models provide exact cosmological solutions~\cite{Akbarieh2026VectorNoether,Akbarieh2026TeleparallelVector}. The method also applies to nonminimally coupled and teleparallel Yang--Mills systems~\cite{Akbarieh2026YangMills,Izadi2026TeleparallelYangMills}. Related scalar--torsion constructions also use symmetry constraints to select coupling functions and integrate the background~\cite{Kucukakca:2013mya,Gecim:2017hmn,Kucukakca:2018twx,Kucukakca2024FTXphi,Dialektopoulos2023Multiscalar}. These are methodological comparisons, since their gravitational and gauge interactions differ from the Abelian coupled-form theory considered here.

The closest scalar--vector precedents are the cosmological analysis of Vakili~\cite{Vakili2014} and the Noether construction of Sun Zhang~\cite{Zhang2015}. The squared combinations of exponentials, their associated scalar--vector coupling relation, and adapted minisuperspace coordinates therefore have clear antecedents. Including a two-form adds another kinetic function and a velocity-linear $BF$ contribution, both of which must enter the determining equations. Our analysis keeps these additional conditions explicit rather than attributing the established scalar--vector structures to the new interaction.

We consider a canonical scalar coupled to three identical Abelian one-form/two-form pairs with scalar-dependent kinetic functions and a constant $BF$ coefficient. Equal-amplitude orthogonal configurations give an exactly isotropic background while retaining the interaction between both form sectors. Our aim is to connect the constrained homogeneous dynamics, perturbative conditions, and symmetry-adapted exact solutions within the same covariant model. Retaining the lapse preserves the Friedmann constraint, and comparisons with uncoupled or reduced-field models use consistent restrictions of the field equations.

For the symmetry analysis, we choose a quadratic effective potential in flat minisuperspace coordinates and solve the autonomous, lapse-preserving Noether conditions with zero boundary function. The regular branches include a de Sitter limit and an exact accelerating family with both form sectors active. We also examine a distinct exponential family with a mixed accelerating attractor and future-bounded normalized linear perturbations at each fixed finite comoving wavenumber. These examples describe potential-supported acceleration modified by form dynamics; they do not constitute an observational fit.

Section~\ref{sec:model} introduces the theory, its isotropic reduction, and consistent limits. Section~\ref{sec:perturbations} develops the perturbation analysis and mixed accelerating solution. Section~\ref{sec:noether} derives the Noether branches and exact cosmologies, Section~\ref{sec:numerics} compares their consistent limits, and Section~\ref{sec:conclusions} summarizes the results.

\section{The model and its homogeneous reduction}
\label{sec:model}

We consider an independent canonical scalar field coupled to three identical Abelian one-form and two-form pairs through scalar-dependent kinetic functions and a constant topological interaction. The triplet structure permits an exactly isotropic configuration, while retaining the interaction between the two form sectors. We first specify the covariant theory and then derive its spatially flat Friedmann--Robertson--Walker (FRW) reduction with the lapse kept arbitrary.

\subsection{Covariant model}
\label{subsec:covariant}

We work in four spacetime dimensions with metric signature $(-,+,+,+)$ and natural units $c=\hbar=1$. The reduced Planck mass is $\Mpl=(8\pi G)^{-1/2}$. For each internal index $r=1,2,3$, the one-form $A^r_\mu$ and the antisymmetric two-form $B^r_{\mu\nu}$ have field strengths
\begin{align}
F^r_{\mu\nu} &= 2\partial_{[\mu}A^r_{\nu]},
&
H^r_{\mu\nu\rho} &= 3\partial_{[\mu}B^r_{\nu\rho]}.
\label{eq:strengths}
\end{align}
The action is
\begin{align}
S ={}& \int \dd^4x\,\sqrt{-g}\left[\frac{\Mpl^2}{2}R-\frac12\partial_\mu\phi\,\partial^\mu\phi-V(\phi)
-\frac{I(\phi)}{4}\sum_{r=1}^{3}F^r_{\mu\nu}F^{r\,\mu\nu}
\right.\nonumber\\[-1pt]
&\left.\hspace{42mm}-\frac{J(\phi)}{12}\sum_{r=1}^{3}H^r_{\mu\nu\rho}H^{r\,\mu\nu\rho}\right]
-\frac{m}{4}\int\dd^4x\,\varepsilon^{\mu\nu\rho\sigma}\sum_{r=1}^{3}B^r_{\mu\nu}F^r_{\rho\sigma}.
\label{eq:action}
\end{align}
Here $m$ is a real constant, $I(\phi)$ and $J(\phi)$ are smooth kinetic functions, and $V(\phi)$ is the scalar potential. The alternating symbol is normalized by $\varepsilon^{0123}=+1$; the corresponding spacetime tensor is $\mathcal{E}^{\mu\nu\rho\sigma}=\varepsilon^{\mu\nu\rho\sigma}/\sqrt{-g}$. In particular, the last integral in Eq.~\eqref{eq:action} uses the symbol and contains no additional metric determinant. For a single one-form/two-form pair, this is the action of Eq.~(2.25) in Ref.~\cite{BeltranAlmeida2019}, with $f_1=I$, $f_2=J$, and $V_{\mathrm{eff}}=V$, after matching the orientation and sign convention for $m$. Equation~\eqref{eq:action} extends this form sector to three identical pairs coupled to the same canonical scalar and metric, admitting an isotropic configuration. The general coupled-form construction and its massive-vector dual are established in Ref.~\cite{BeltranAlmeida2020}.

The Abelian gauge transformations are
\begin{equation}
A^r_\mu\longrightarrow A^r_\mu+\partial_\mu\alpha^r,
\qquad
B^r_{\mu\nu}\longrightarrow B^r_{\mu\nu}+2\partial_{[\mu}\Lambda^r_{\nu]}.
\label{eq:gauge}
\end{equation}
For constant $m$, the two-form variation of the $BF$ term is a boundary term, using $\partial_{[\mu}F^r_{\nu\rho]}=0$. Thus the action respects these transformations under the usual boundary conditions. We also define the composite quantity
\begin{equation}
\xi\equiv\frac12\sum_{r=1}^{3}B^r_{\mu\nu}B^{r\,\mu\nu}.
\label{eq:xi-cov}
\end{equation}
Although $\xi$ is a spacetime scalar, it is not invariant under the two-form gauge transformation in Eq.~\eqref{eq:gauge}. It is used only to describe the amplitude of a specified field configuration. It is not an independent scalar field, and the functions $I$, $J$, and $V$ in Eq.~\eqref{eq:action} depend on $\phi$, not on $\xi$.

Variation with respect to $\phi$, $A^r_\nu$, and $B^r_{\mu\nu}$ gives, respectively,
\begin{subequations}\label{eq:cov-eom}
\begin{align}
\Box\phi-V_{,\phi}-\frac{I_{,\phi}}{4}\sum_r F^r_{\mu\nu}F^{r\,\mu\nu}
-\frac{J_{,\phi}}{12}\sum_r H^r_{\mu\nu\rho}H^{r\,\mu\nu\rho} &= 0,
\label{eq:cov-scalar}\\
\nabla_\mu\!\left(I F^{r\,\mu\nu}\right)-\frac{m}{6}\mathcal{E}^{\nu\alpha\beta\gamma}H^r_{\alpha\beta\gamma} &= 0,
\label{eq:cov-A}\\
\nabla_\rho\!\left(J H^{r\,\rho\mu\nu}\right)-\frac{m}{2}\mathcal{E}^{\mu\nu\rho\sigma}F^r_{\rho\sigma} &= 0,
\label{eq:cov-B}
\end{align}
\end{subequations}
where $\Box=\nabla_\mu\nabla^\mu$ and a comma denotes differentiation with respect to $\phi$. Metric variation yields $\Mpl^2G_{\mu\nu}=T_{\mu\nu}$. Since the $BF$ term is metric independent, its direct stress tensor vanishes identically, $T^{BF}_{\mu\nu}=0$. Nevertheless, it changes the evolution of both form sectors through Eqs.~\eqref{eq:cov-A} and~\eqref{eq:cov-B}.

We restrict attention to $I(\phi)>0$ and $J(\phi)>0$ along the background, so that both form kinetic terms have the standard signs. Section~\ref{sec:perturbations} derives the full quadratic actions and sufficient conditions for a positive reduced kinetic matrix, and applies them to an accelerating solution. For $m\neq0$, a one-form/two-form pair has a local St\"uckelberg massive-vector description, with the algebraic mass scale $\mu^2(\phi)=m^2/[I(\phi)J(\phi)]$ after freezing the kinetic coefficients~\cite{Cremmer:1973mg,Allen:1990gb,BeltranAlmeida2020}. Time-dependent normalization introduces additional terms in the mode equations, so this scale alone does not specify the full perturbation frequency. The coupled pairs therefore cannot be identified with two ordinary scalar fields, and the canonical field $\phi$ remains an independent part of the theory.

\subsection{Isotropic FRW configuration}
\label{subsec:frw}

The spatially flat FRW line element is written as
\begin{equation}
\dd s^2=-N^2(t)\,\dd t^2+a^2(t)\delta_{ij}\,\dd x^i\dd x^j,
\qquad
\phi=\phi(t),
\qquad
\Hc\equiv\frac{\dot a}{Na},
\label{eq:metric}
\end{equation}
where $N(t)>0$ is the lapse, $a(t)>0$ is the scale factor, and a dot denotes differentiation with respect to $t$. The notation $\Hc$ distinguishes the Hubble expansion rate from the three-form field strengths $H^r_{\mu\nu\rho}$.

We choose equal-amplitude orthogonal triplets~\cite{Yamamoto:2012sq,Chen:2022ccf},
\begin{equation}
A^r_0=0,
\qquad
A^r_i=u(t)\delta_i^r,
\qquad
B^r_{0i}=0,
\qquad
B^r_{ij}=v(t)\epsilon^r{}_{ij},
\label{eq:ansatz}
\end{equation}
with $\epsilon^r{}_{ij}=\delta^{rk}\epsilon_{kij}$ and $\epsilon_{123}=+1$. The two-form potential is purely spatial, often called a magnetic-type configuration. The common amplitudes and identical couplings make the summed stress tensor isotropic: a spatial rotation can be compensated by the same internal rotation of the triplets. In particular, $\sum_r\delta_i^r\delta_j^r=\delta_{ij}$ and $\sum_{r,k}\epsilon^r{}_{ik}\epsilon^r{}_{jk}=2\delta_{ij}$. No average over field directions is required.

The nonvanishing field-strength components are
\begin{equation}
F^r_{0i}=\dot u\,\delta_i^r,
\qquad
H^r_{0ij}=\dot v\,\epsilon^r{}_{ij},
\qquad
F^r_{ij}=0,
\qquad
H^r_{ijk}=0.
\label{eq:background-strengths}
\end{equation}
The temporal equations associated with $A^r_0$ and $B^r_{0i}$ are satisfied by this homogeneous configuration: their spatial divergences vanish, as do the corresponding $H^r_{ijk}$ and $F^r_{ij}$ sources. The contractions entering the action reduce to
\begin{align}
\sum_r F^r_{\mu\nu}F^{r\,\mu\nu} &=-\frac{6\dot u^2}{N^2a^2},
&
\sum_r H^r_{\mu\nu\rho}H^{r\,\mu\nu\rho} &=-\frac{18\dot v^2}{N^2a^4},
\label{eq:kinetic-contractions}\\
\varepsilon^{\mu\nu\rho\sigma}\sum_r B^r_{\mu\nu}F^r_{\rho\sigma} &=12v\dot u.
\label{eq:bf-contraction}
\end{align}
The numerical factors include the three internal components and all antisymmetric index permutations. Equation~\eqref{eq:bf-contraction} fixes the sign of the reduced topological coupling for the orientation chosen above.

On this configuration, the composite in Eq.~\eqref{eq:xi-cov} becomes
\begin{equation}
\xi=\frac{3v^2}{a^4}=3\psi^2,
\qquad
\psi\equiv\frac{v}{a^2},
\qquad
\dot v=a^2\!\left(\dot\psi+2\frac{\dot a}{a}\psi\right).
\label{eq:xi-frw}
\end{equation}
Here $\psi$ is a rescaled two-form amplitude. Neither $\psi$ nor $\xi$ introduces an additional independent matter degree of freedom. In particular, $\xi$ loses the sign of $v$ and does not provide a globally equivalent replacement for $v$. The reduced variables $u(t)$ and $v(t)$ should likewise be interpreted as form amplitudes. Even in the decoupled limit $m=0$, dualizing this $H^r_{0ij}$ configuration gives scalar fields with spatial gradients, rather than homogeneous canonical scalars, as expected from cosmological two-form duality~\cite{Aoki:2022ylc}.

\subsection{Pointlike Lagrangian and background equations}
\label{subsec:pointlike}

We retain the lapse when reducing the action, since premature restrictions on metric variables can discard field equations or constraints~\cite{Frausto2025SymmetryReduction}. We integrate over a unit comoving spatial volume and remove the usual Einstein--Hilbert total derivative, or equivalently include the appropriate gravitational boundary term. For the metric in Eq.~\eqref{eq:metric}, the gravitational contribution obeys
\begin{equation}
\frac{\Mpl^2}{2}Na^3R
=-\frac{3\Mpl^2a\dot a^2}{N}
+\frac{\dd}{\dd t}\!\left(\frac{3\Mpl^2a^2\dot a}{N}\right).
\label{eq:eh-reduction}
\end{equation}
Using Eqs.~\eqref{eq:kinetic-contractions} and~\eqref{eq:bf-contraction}, the reduced action $S=\int\dd t\,L$ therefore has the pointlike Lagrangian
\begin{equation}
\boxed{\displaystyle
L=-\frac{3\Mpl^2a\dot a^2}{N}
+\frac{a^3\dot\phi^2}{2N}
+\frac{3aI(\phi)\dot u^2}{2N}
+\frac{3J(\phi)\dot v^2}{2aN}
-3mv\dot u-Na^3V(\phi).}
\label{eq:lagrangian}
\end{equation}
Its configuration variables are $(a,\phi,u,v)$, with $N$ acting as a Lagrange multiplier. The term $-3mv\dot u$ contains neither $a$ nor $N$, as expected from the metric independence of the covariant interaction. Integration by parts would replace it with $3mu\dot v$ without changing the equations of motion.

The lapse must be varied before choosing a time coordinate. Its equation gives the Friedmann constraint
\begin{equation}
3\Mpl^2\Hc^2
=\frac{\dot\phi^2}{2N^2}+V(\phi)+\rho_F+\rho_H,
\qquad
\rho_F\equiv\frac{3I\dot u^2}{2N^2a^2},
\quad
\rho_H\equiv\frac{3J\dot v^2}{2N^2a^4}.
\label{eq:friedmann}
\end{equation}
The quantities $\rho_F$ and $\rho_H$ are the total kinetic energy densities of the one-form and two-form triplets, respectively. Variation with respect to the three matter amplitudes yields
\begin{subequations}\label{eq:background-eom}
\begin{align}
\frac{\dd}{\dd t}\!\left(\frac{a^3\dot\phi}{N}\right)
+Na^3V_{,\phi}
-\frac{3aI_{,\phi}\dot u^2}{2N}
-\frac{3J_{,\phi}\dot v^2}{2aN} &=0,
\label{eq:background-scalar}\\
\frac{\dd}{\dd t}\!\left(\frac{aI\dot u}{N}\right)-m\dot v &=0,
\label{eq:background-u}\\
\frac{\dd}{\dd t}\!\left(\frac{J\dot v}{aN}\right)+m\dot u &=0.
\label{eq:background-v}
\end{align}
\end{subequations}
These equations agree with the direct substitution of the ansatz into Eqs.~\eqref{eq:cov-eom}. Cosmic time corresponds to $N=1$ and may be imposed after obtaining the constraint and evolution equations.

The total isotropic energy density and pressure are
\begin{equation}
\rho=\frac{\dot\phi^2}{2N^2}+V+\rho_F+\rho_H,
\qquad
p=\frac{\dot\phi^2}{2N^2}-V+\frac{\rho_F}{3}-\frac{\rho_H}{3}.
\label{eq:rho-pressure}
\end{equation}
Thus the kinetic one-form and two-form contributions have instantaneous pressure-to-density ratios $1/3$ and $-1/3$, respectively, although their energy densities need not follow the uncoupled dilution laws. Variation with respect to $a$, together with Eq.~\eqref{eq:friedmann}, gives
\begin{equation}
-\frac{2\Mpl^2}{N}\dot\Hc
=\rho+p
=\frac{\dot\phi^2}{N^2}+\frac{4}{3}\rho_F+\frac{2}{3}\rho_H.
\label{eq:raychaudhuri}
\end{equation}
Equations~\eqref{eq:background-eom} also imply $\dot\rho+3N\Hc(\rho+p)=0$, providing a consistency check on the reduction. The acceleration equation can be written as
\begin{equation}
\frac{1}{N}\dot\Hc+\Hc^2
=\frac{\ddot a}{N^2a}-\frac{\dot a\dot N}{N^3a}
=\frac{1}{3\Mpl^2}\!\left(V-\frac{\dot\phi^2}{N^2}-\rho_F\right).
\label{eq:acceleration}
\end{equation}
For positive kinetic functions, the form sector alone satisfies $\rho_{\mathrm{forms}}+3p_{\mathrm{forms}}=2\rho_F\geq0$. The $BF$ interaction can modify the energy exchange and background evolution, but it supplies no direct negative-pressure source. Accelerated expansion in this model consequently requires the potential to satisfy $V>\dot\phi^2/N^2+\rho_F$. The functions $I(\phi)$, $J(\phi)$, and $V(\phi)$ remain unspecified at this stage.

\subsection{Consistent limits and relation to established models}
\label{subsec:known-limits}

Recent form cosmologies include scalar--vector dark-energy attractors~\cite{Orjuela-Quintana:2021zoe}, interacting vector-like dark energy~\cite{Rodriguez-Benites:2023otm}, antisymmetric-tensor inflation~\cite{Aashish:2021gdf}, and scalar--two-form power-law solutions~\cite{Pham:2023evo}. Two-form interactions with cold dark matter provide another route to late-time fixed points~\cite{Orjuela-Quintana:2022jrg}. These theories use different potentials, matter couplings, or background symmetries, so they are contextual comparisons rather than automatic limits of the present action. Several familiar theories do follow from Eq.~\eqref{eq:action} by consistent restrictions. First, setting all form field strengths to zero gives minimally coupled canonical scalar cosmology, independently of $m$. This includes canonical quintessence models~\cite{Andriot:2024jsh}; for an exponential potential it contains the scalar-field solutions studied in Ref.~\cite{Copeland1998}. This is a restriction of the field configuration, rather than a replacement of the independent scalar $\phi$ by a form amplitude.

Second, when $m=0$, either form sector may be removed consistently. The restriction $H^r_{\mu\nu\rho}=0$ gives a scalar coupled to an Abelian Maxwell triplet, with the identification $I(\phi)=f^2(\phi)$ used in Ref.~\cite{Yamamoto:2012sq}. Retaining only one nonzero vector gives the covariant scalar--Maxwell model of Ref.~\cite{Watanabe:2009ct}, although a single nonzero background vector generally requires an anisotropic metric. Likewise, $m=0$ and $F^r_{\mu\nu}=0$ give scalar--two-form cosmology. Its single-two-form version has $J=f^2$ in Ref.~\cite{Ohashi:2013mka}; the orthogonal triplet has $J=g^2$ and the same spatial configuration as the two-form sector of Ref.~\cite{Chen:2022ccf}. Thus both the field content and the background symmetry must be specified when comparing these limits. Isotropic one-form and two-form triplets have already been studied separately without the present $BF$ coupling~\cite{Chen:2022ccf}, and the Noether-symmetry method has been applied to a cosmic vector triad~\cite{Zhang:2009yu}. The calculations below concern the simultaneously $BF$-coupled isotropic three-pair system and its background and perturbation dynamics.

In the homogeneous decoupled system the first integrals are
\begin{equation}
 \frac{aI\dot u}{N}=q_F,\qquad
 \frac{J\dot v}{aN}=q_H,\qquad
 \rho_F=\frac{3q_F^2}{2Ia^4},\qquad
 \rho_H=\frac{3q_H^2}{2Ja^2},
 \label{eq:decoupled-first-integrals}
\end{equation}
where $q_F$ and $q_H$ are constants. In particular, constant $I$ and $J$ give the radiation-like and gradient-like dilution laws $a^{-4}$ and $a^{-2}$. The massless two-form admits a local scalar dual with kinetic term $-(\partial\chi^r)^2/(2J)$; for the background in Eq.~\eqref{eq:ansatz}, its scalar representative is $\chi^r=q_H x^r+\mathrm{constant}$, with the orientation of Eq.~\eqref{eq:action}. It is therefore not a second homogeneous quintessence field.

For $m\ne0$, retaining only one nontrivial form sector is generally inconsistent: the equation of the omitted sector would also force the retained field strength to vanish. Instead, each coupled pair has a local massive-vector description, the topological mass-generation mechanism of Refs.~\cite{Cremmer:1973mg,Allen:1990gb}. With constant positive coefficients, $A_c^r=\sqrt I A^r$ and $B_c^r=\sqrt J B^r$ bring its action to the standard kinetic normalization, and the mass is $\mu=|m|/\sqrt{IJ}$. For variable coefficients the local dual action is still available, but time-dependent normalizations affect the perturbation equations~\cite{BeltranAlmeida2020}; this issue is treated in Section~\ref{sec:perturbations}.

These identifications do not encompass theories with nonminimal curvature couplings such as $f(\xi)R$, torsion couplings, additional parity-odd kinetic interactions, or non-Abelian Yang--Mills self-interactions. None of those operators occurs in Eq.~\eqref{eq:action}.

\section{Perturbative stability and degrees of freedom}
\label{sec:perturbations}

\subsection{Local duality and constraint counting}
\label{subsec:duality-dof}

The kinetic signs and the number of propagating fields can be established without assuming homogeneity. Introduce
\begin{equation}
 h^{r\mu}=\frac16\mathcal E^{\mu\nu\rho\sigma}H^r_{\nu\rho\sigma},
 \qquad H^r_{\mu\nu\rho}=\mathcal E_{\mu\nu\rho\sigma}h^{r\sigma}.
\end{equation}
After a boundary term has been removed, the first-order form Lagrangian for one pair is
\begin{equation}
 \mathcal L_{\rm first}=-\frac I4F_{\mu\nu}F^{\mu\nu}
 +\frac J2h_\mu h^\mu-h^\mu(mA_\mu-\partial_\mu\chi).
\end{equation}
Variation of $\chi$ enforces $\nabla_\mu h^\mu=0$, locally recovering the two-form, whereas eliminating $h^\mu$ yields
\begin{equation}
 {\mathcal L_{\rm dual}=-\frac I4F_{\mu\nu}F^{\mu\nu}
 -\frac1{2J}(\partial_\mu\chi-mA_\mu)(\partial^\mu\chi-mA^\mu).}
 \label{eq:dual-stueckelberg}
\end{equation}
This is invariant under $A_\mu\mapsto A_\mu+\partial_\mu\alpha$ and $\chi\mapsto\chi+m\alpha$. For $m\ne0$, the gauge-invariant combination $C_\mu=A_\mu-m^{-1}\partial_\mu\chi$ has the Proca Lagrangian
\begin{equation}
 \mathcal L_{\rm P}=-\frac I4F_{\mu\nu}(C)F^{\mu\nu}(C)-\frac q2C_\mu C^\mu,
 \qquad q\equiv\frac{m^2}{J}>0,
 \qquad \mu^2\equiv\frac qI.
 \label{eq:proca-dual}
\end{equation}
The equivalence is local and does not identify global flux sectors. Recent canonical analyses of massive two-form/vector duality emphasize that the interaction terms must also be mapped when comparing theories~\cite{Barbosa:2022zfm,Hell:2021wzm}; agreement of the free degree count alone is insufficient. Quantum effective actions and topology-dependent duality effects require separate treatment~\cite{Kuzenko:2020zad,Canete:2026vxk}, whereas the representation used here is classical and local. It is the specialization of the topological mass mechanism discussed in Refs.~\cite{Cremmer:1973mg,Allen:1990gb,BeltranAlmeida2020}. At $m=0$ one instead has a Maxwell field and a dual scalar with kinetic coefficient $J^{-1}$; taking a unitary-gauge formula through $m=0$ is inappropriate.

For completeness, the constraint count can also be performed before dualization. On an ADM spatial slice, set $b^i=\epsilon^{ijk}B_{jk}/2$ and use momenta $\pi^i$ and $p_i$ conjugate to $A_i$ and $b^i$, respectively. The lapse Hamiltonian density for one pair is
\begin{align}
 \mathcal H_{\perp,\mathrm{pair}}={}&
 \frac{h_{ij}(\pi^i+mb^i)(\pi^j+mb^j)}{2\sqrt h\,I}
 +\frac{\sqrt h\,h^{ij}p_i p_j}{2J}
 +\frac{\sqrt h\,I}{4}F_{ij}F^{ij}
 +\frac{J}{2\sqrt h}(\partial_i b^i)^2.
 \label{eq:bf-hamiltonian}
\end{align}
Here $h=\det h_{ij}$ and $\epsilon^{ijk}$ is the alternating symbol, so $b^i$ is a spatial vector density. The multipliers $A_0$ and $B_{0i}$ enforce
\begin{equation}
 \mathcal G=\partial_i\pi^i=0,
 \qquad
 \mathcal C^i=\epsilon^{ijk}\partial_j(p_k+mA_k)=0,
 \qquad \partial_i\mathcal C^i\equiv0.
 \label{eq:bf-constraints}
\end{equation}
These constraints are first class; the identity in the last expression leaves only two independent components of $\mathcal C^i$. Consequently the six spatial configuration variables of each pair carry $6-1-2=3$ local physical degrees of freedom. Equivalently, including the temporal potentials gives four primary and three independent secondary first-class constraints on ten configuration variables. Adding the scalar and gravity gives
\begin{equation}
 {N_{\rm physical}=2+1+3\times3=12.}
 \label{eq:physical-dof}
\end{equation}
The same count holds at $m=0$, with two Maxwell and one two-form polarization per pair. Smooth scalar-dependent $I$ and $J$ do not change the constraint rank on the domain $I,J>0$. A constant $m$ is required for the two-form gauge invariance used in this argument.

\subsection{Perturbations under the diagonal rotation symmetry}
\label{subsec:svt}

For the remainder of this section $m\ne0$, and cosmic time is used, $N=1$ on the background. The gauge-invariant Proca field dual to Eq.~\eqref{eq:ansatz} is
\begin{equation}
 C_0^r=0,\qquad C_i^r=\mathcal{A}\delta_i^r,\qquad
 \mathcal{A}=u-\frac{c_H}{m}=-\frac{J\dot v}{am},\qquad
 c_H=mu+\frac{J\dot v}{a}=\mathrm{constant}.
 \label{eq:dual-background}
\end{equation}
Thus $\dot{\mathcal{A}}=\dot u$ and $3q\mathcal{A}^2/(2a^2)=\rho_H$; the constant in $u$ is absorbed into the background St\"uckelberg gradient.

The unbroken simultaneous spatial and internal rotations permit a scalar--vector--tensor decomposition of the form perturbations~\cite{Yamamoto:2012sq,Chen:2022ccf}. Exact background isotropy does not remove the gauge-sector signatures in fluctuations, as recent vector-inflation bispectrum calculations illustrate~\cite{Chen:2026rwr}; the present analysis remains at quadratic order. Align a nonzero Fourier wavevector with the $z$ direction, and use spatially flat scalar and vector metric gauges, retaining the transverse traceless metric tensor $\gamma_{ij}$. The matrix $\delta C_i{}^r$ contributes three helicity-zero amplitudes, four helicity-one amplitudes, and two helicity-two amplitudes. Together with $\delta\phi$ and the two metric tensor polarizations, the sectors contain
\begin{equation}
 N_S=4,\qquad N_V=4,\qquad N_T=4,\qquad N_S+N_V+N_T=12.
\end{equation}
One helicity-zero amplitude is antisymmetric in the spatial--internal perturbation matrix. The internal rotation symmetry is global and does not remove this physical orientation mode. Below, every action is quadratic and exact in $k>0$, with a common positive Fourier normalization suppressed. The lapse, shift, and temporal vectors are eliminated explicitly; the strictly homogeneous sector is considered separately in the final subsection.

\subsection{Tensor and transverse vector sectors}
\label{subsec:tensor-vector}

\paragraph{Tensor modes.}
The tensor perturbations are defined by $h_{ij}=a^2(e^\gamma)_{ij}$ and $\delta C_i{}^r=T_{ir}$, where both $\gamma$ and $T$ are symmetric, transverse and traceless. For a real Fourier mode (or with products interpreted as Hermitian products),
\begin{align}
S_T^{(2)}=\int \dd t\,\dd^3k\,\bigg\{&\frac{a^3\Mpl^2}{8}\left(\dot\gamma_{ij}\dot\gamma_{ij}-\frac{k^2}{a^2}\gamma_{ij}\gamma_{ij}\right)
+\frac{aI}{2}\left(\dot T_{ij}\dot T_{ij}-2\dot{\mathcal{A}}\gamma_{ij}\dot T_{ij}+\frac{\dot{\mathcal{A}}^2}{2}\gamma_{ij}\gamma_{ij}\right)\nonumber\\
&-\frac{Ik^2}{2a}T_{ij}T_{ij}-\frac{aq}{2}\left(T_{ij}T_{ij}-2\mathcal{A}\gamma_{ij}T_{ij}+\frac{\mathcal{A}^2}{2}\gamma_{ij}\gamma_{ij}\right)\bigg\}.
\end{align}
The tensor lapse, shift and temporal-vector perturbations vanish by their representation content. This is an unredefined, exact quadratic action, and it retains the one-derivative mixing and the background-dependent mass terms. Its two kinetic blocks are strictly positive for $\Mpl^2>0$ and $I>0$. There are four propagating tensor modes, and their two high-frequency characteristic speeds are unity. The one-derivative mixing and effective frequency terms are retained explicitly for subsequent mode evolution.

\paragraph{Transverse vector sector.}
Take $\mathbf k=k\hat{\mathbf z}$ and use the spatial vector diffeomorphism to set the vector perturbation of $h_{ij}$ to zero. For each $\alpha=x,y$, define $\delta C_\alpha{}^3=X\cos kz$, $\delta C_3{}^\alpha=-Y\cos kz$, $C_0{}^\alpha=Z\sin kz$, and $N^\alpha=S\sin kz$. An irrelevant common Fourier normalization is suppressed. The action before eliminating constraints is
\begin{align}
L_V^{(2)}={}&\frac{aI}{2}\left[\dot X^2+(\dot Y+kZ)^2\right]
+\frac{a^3q}{2}(Z-\mathcal{A} S)^2+\frac{a^3\Mpl^2k^2}{4}S^2-aI\dot{\mathcal{A}}kSX\nonumber\\
&-\frac12\left(aq+\frac{Ik^2}{a}\right)X^2-\frac{aq}{2}Y^2.
\end{align}
The nondynamical pair obeys
\begin{equation}
\begin{pmatrix}aIk^2+a^3q&-a^3q\mathcal{A}\\-a^3q\mathcal{A}&a^3q\mathcal{A}^2+a^3\Mpl^2k^2/2\end{pmatrix}\begin{pmatrix}Z\\S\end{pmatrix}
=-\begin{pmatrix}aIk\dot Y\\-aI\dot{\mathcal{A}}kX\end{pmatrix}.
\end{equation}
For $k>0$, $I,q,\Mpl^2>0$, this matrix is nonsingular. Define
\begin{equation}
D_V=Ik^2+a^2q+\frac{2Iq\mathcal{A}^2}{\Mpl^2},\qquad
R_V=\frac{2aI^2q\mathcal{A}\dot{\mathcal{A}}}{\Mpl^2D_V}.
\end{equation}
The fully reduced action is
\begin{align}
L_{V,\mathrm{red}}^{(2)}={}&\frac{aI}{2}\dot X^2+\frac{a^3Iq}{2D_V}\dot Y^2+R_VX\dot Y
-\frac{\mathcal M_X}{2}X^2-\frac{aq}{2}Y^2,\\
\mathcal M_X={}&aq+\frac{Ik^2}{a}+\frac{2I^2\dot{\mathcal{A}}^2(Ik^2+a^2q)}{a\Mpl^2D_V}.
\end{align}
This demonstrates a positive all-$k$ kinetic matrix without neglecting gravitational momentum constraints. Each transverse helicity carries two modes, giving four vector-sector degrees of freedom. The exact time-kinetic normalization of the longitudinal amplitude is
\begin{equation}
 Y_{\mathrm{can}}=\sqrt{\frac{a^3Iq}{D_V}}\,Y,\qquad
 \frac{D_V}{Ia^2}=\frac{k^2}{a^2}+\mu^2+\frac{4\rho_H}{3\Mpl^2}.
 \label{eq:vector-canonical-normalization}
\end{equation}
Here $\rho_H=3q\mathcal{A}^2/(2a^2)$. Thus $Y_{\mathrm{can}}\simeq\sqrt{a^3q}\,Y/k$ requires
\begin{equation}
 \frac{k^2}{a^2}\gg\mu^2+\frac{4\rho_H}{3\Mpl^2}.
 \label{eq:vector-normalization-limit}
\end{equation}
The gravitational scale $\sqrt{4\rho_H/(3\Mpl^2)}$ need not be controlled by $\Hc$ on the general domain $V+\rho_H>0$, which permits negative $V$. Equation~\eqref{eq:vector-normalization-limit} specifies the normalization approximation; an adiabatic mode approximation additionally requires control of background variation and the remaining mixing terms. The luminal principal characteristics are established independently in Section~\ref{subsec:characteristics}. The retained time dependence and $R_V$ determine the finite-wavelength evolution studied below.

\subsection{Scalar constraints and the reduced quadratic action}
\label{subsec:scalar-action}

For a nonzero wavevector aligned with $z$, use spatially flat gauge, $h_{ij}=a^2\delta_{ij}$ in the scalar sector. The following real standing-wave convention avoids ambiguous Fourier factors of $i$:
\begin{equation}
 \delta\phi=f\cos kz,\quad N=1+n\cos kz,\quad N^z=s\sin kz,
 \quad \delta C_0^z=A\sin kz,
\end{equation}
\begin{equation}
 \delta C_i^r=\begin{pmatrix}T/\sqrt2&P/\sqrt2&0\\-P/\sqrt2&T/\sqrt2&0\\0&0&L\end{pmatrix}\cos kz.
\end{equation}
The common positive spatial averaging factor is suppressed. The amplitudes $(f,T,L)$ form the coupled scalar block; $P$ is the antisymmetric helicity-zero mode. These labels refer to the rotation decomposition and do not assign intrinsic spatial parity to the original form fields. No metric scalar propagates separately in this gauge. There are four physical scalar-sector degrees of freedom.

Define
\begin{equation}
 g=\frac I{a^2},\quad d=\dot{\mathcal{A}},\quad \varpi=\dot\phi,\quad h=2\Mpl^2\Hc,
 \quad \mathcal{U}=V+\frac{3q\mathcal{A}^2}{2a^2},\quad r=\frac{I_{,\phi}d}{a^2},
\end{equation}
\begin{equation}
 B=V_{,\phi}+\frac{3q_{,\phi}\mathcal{A}^2}{2a^2}+\frac{3I_{,\phi}d^2}{2a^2},
 \qquad \mathcal{M}_f^2=V_{,\phi\phi}+\frac{3q_{,\phi\phi}\mathcal{A}^2}{2a^2}-\frac{3I_{,\phi\phi}d^2}{2a^2}.
\end{equation}
The Friedmann constraint is used only to replace the coefficient of $n^2$ by $-\mathcal{U}$.

\paragraph{Exact unreduced quadratic scalar action.}
With $S^{(2)}=\int \dd t\,a^3\mathcal L_2$, the complete scalar action, including the decoupled antisymmetric amplitude, is
\begin{align}
\label{eq:scalar-full-action}
\mathcal L_2={}&\frac12\dot f^2-\frac12\left(\frac{k^2}{a^2}+\mathcal{M}_f^2\right)f^2
 +\frac g2\left[\dot T^2+(\dot L-kA)^2+\dot P^2-\frac{k^2}{a^2}(T^2+P^2)\right]
\\
&-\frac q{2a^2}(T^2+L^2+P^2)\\
&+rf(\sqrt2\dot T+\dot L-kA)-\frac{q_{,\phi}\mathcal{A}}{a^2}f(\sqrt2T+L)\\
&-n\left[\varpi\dot f+gd(\sqrt2\dot T+\dot L-kA)+Bf+\frac{q\mathcal{A}}{a^2}(\sqrt2T+L)\right]
\\
&-\mathcal{U}n^2-hkns+ks(\varpi f+\sqrt2gdT)+\frac q2(A-\mathcal{A}s)^2.
\end{align}
The $ks\sqrt2gdT$ term is the electric-field contribution from $F_{0i}-N^jF_{ji}$.

\paragraph{All three scalar constraints and their exact solution.}
The lapse, shift, and temporal-vector equations are
\begin{align}
\label{eq:scalar-constraints}
0={}&-2\mathcal{U}n-hks+gdkA-\varpi\dot f-gd(\sqrt2\dot T+\dot L)-Bf-\frac{q\mathcal{A}}{a^2}(\sqrt2T+L),\\
0={}&-hkn-q\mathcal{A}(A-\mathcal{A}s)+k(\varpi f+\sqrt2gdT),\\
0={}&(gk^2+q)A-q\mathcal{A}s+gdk n-gk\dot L-rkf.
\end{align}
Equivalently, with $z=(n,s,A)^T$,
\begin{equation}
\label{eq:scalar-constraint-matrix}
 \mathsf C=\begin{pmatrix}-2\mathcal{U}&-hk&gdk\\-hk&q\mathcal{A}^2&-q\mathcal{A}\\gdk&-q\mathcal{A}&gk^2+q\end{pmatrix},
\end{equation} 
\begin{equation}
\label{eq:scalar-constraint-source}
 j=\begin{pmatrix}-\varpi\dot f-gd(\sqrt2\dot T+\dot L)-Bf-\frac{q\mathcal{A}}{a^2}(\sqrt2T+L)\\k(\varpi f+\sqrt2gdT)\\-gk\dot L-rkf\end{pmatrix},
 \qquad z=-\mathsf C^{-1}j.
\end{equation}
Let $\mathcal L_0=\mathcal L_2|_{n=s=A=0}$. Then the completely reduced, exact finite-$k$ action is the explicit Schur complement
\begin{equation}
\label{eq:scalar-schur}
 {\mathcal L_{\mathrm{red}}=\mathcal L_0-\frac12j^T\mathsf C^{-1}j.}
\end{equation}
All entries of the constraint matrix and source vector are specified above, so this reduced action retains the full mass, derivative-mixing, and finite-wavelength dependence.
\begin{equation}
 \det \mathsf C=-k^2\Delta,\qquad
 \Delta=gh^2k^2+q(h-gd\mathcal{A})^2+2\mathcal{U} gq\mathcal{A}^2.
\end{equation}
Thus the constraint inversion has no finite-$k$ pole for $\mathcal{U}>0$, $g,q>0$, $\Hc\ne0$. The exactly homogeneous mode is to be treated before dividing by $k$; scalar/vector/tensor helicity decomposition and longitudinal shift gauge fixing are singular at $k=0$.

\paragraph{Exact kinetic matrix.}
For $X=(f,T,L)^T$ the coefficient of $\frac12\dot X^TK\dot X$ is
\begin{equation}
\label{eq:scalar-kinetic-matrix}
 {K=\operatorname{diag}(1,g,0)+\frac{gq}{\Delta}
 \begin{pmatrix}\varpi\mathcal{A}\\\sqrt2gd\mathcal{A}\\h\end{pmatrix}
 \begin{pmatrix}\varpi\mathcal{A}&\sqrt2gd\mathcal{A}&h\end{pmatrix}.}
\end{equation}
Equivalently,
\begin{equation}
 2\mathcal L_{\rm kinetic}=\dot f^2+g\dot T^2+
 \frac{gq}{\Delta}(\varpi\mathcal{A}\dot f+\sqrt2gd\mathcal{A}\dot T+h\dot L)^2+g\dot P^2.
\end{equation}
The three leading Sylvester minors are
\begin{equation}
\label{eq:scalar-sylvester}
 1+\frac{g\varpi^2q\mathcal{A}^2}{\Delta},\qquad
 g\left[1+\frac{gq\mathcal{A}^2(\varpi^2+2gd^2)}{\Delta}\right],\qquad
 \det K=\frac{h^2g^2q}{\Delta}.
\end{equation}
All are strictly positive under the stated conditions. The scalar kinetic matrix is therefore positive at every finite wavelength. The apparent $K_{LL}\sim k^{-2}$ at large $k$ is the usual Proca longitudinal normalization; using the St\"uckelberg scalar makes its ordinary positive kinetic term manifest.

\subsection{Healthy domain and characteristic propagation}
\label{subsec:characteristics}

Combining the tensor, vector, and scalar reductions gives the sufficient domain
\begin{equation}
 \Mpl^2>0,\qquad I>0,\qquad J>0,\qquad m\ne0,\qquad
 \Hc\ne0,\qquad \mathcal{U}=V+\rho_H>0.
 \label{eq:sufficient-no-ghost}
\end{equation}
On this domain all twelve physical modes have a positive-definite kinetic matrix at every finite $k>0$, and the auxiliary-field reductions have no finite-$k$ pole. In the scalar sector this follows directly from $\Delta>0$ and the displayed Sylvester minors; in the vector sector it follows from $D_V>0$; the tensor kinetic blocks are manifestly positive. These are sufficient conditions, and backgrounds outside this domain can be studied by retaining their own constraint equations. The massless limit is regular in the original form variables, whose degree-of-freedom count was given above, rather than in the unitary Proca coordinates.

\paragraph{Direct positive-square proof.}
The same kinetic result can be verified directly before decomposing the form perturbations into helicities. Write $N=1+n$, $N^i=s^i$, $C_0^r=\theta^r$, $C_i^r=\mathcal{A}\delta_i^r+\delta C_i^r$, and set $v_\phi=\delta\dot\phi$, $v_i^r=\delta\dot C_i^r$. In spatially flat gauge with the metric tensor $\gamma_{ij}$ retained, the complete part of the quadratic density that determines its velocity Hessian is
\begin{align}
 \frac{\mathcal L_{\mathrm{Hess}}^{(2)}}{a^3}={}&
 \frac{\Mpl^2}{8}\dot\gamma_{ij}\dot\gamma_{ij}
 +\frac12v_\phi^2+\frac{I}{2a^2}\sum_r(v_i^r-\partial_i\theta^r)^2
 +\frac q2\sum_r(\theta^r-\mathcal{A}s^r)^2\nonumber\\
 &+\frac{\Mpl^2}{4}(\partial_i s_T^j)^2
 -\mathcal{U}n^2-n\mathcal R,\label{eq:full-velocity-hessian}\\
 \mathcal R={}&\dot\phi\,v_\phi
 +\frac{I\dot{\mathcal{A}}}{a^2}\sum_r(v_r^r-\partial_r\theta^r)
 +2\Mpl^2\Hc\partial_i s^i.\label{eq:kinetic-R}
\end{align}
Here $s_T^i$ is the transverse shift, and spatial integrations by parts are understood with periodic or decaying perturbations. Terms containing an undifferentiated physical perturbation do not contribute to this Hessian. The background constraint fixes the lapse coefficient through
\begin{equation}
 -3\Mpl^2\Hc^2+\frac{\dot\phi^2}{2}
 +\frac{3I\dot{\mathcal{A}}^2}{2a^2}=-\mathcal{U}.
\end{equation}
Eliminating $n=-\mathcal R/(2\mathcal{U})$ replaces its two terms by $+\mathcal R^2/(4\mathcal{U})$. The remaining expression is a sum of positive squares on Eq.~\eqref{eq:sufficient-no-ghost}. Eliminating $\theta^r$ and $s^i$ minimizes that expression over the auxiliary variables, so the reduced kinetic form is nonnegative.

Strict positivity at finite $k>0$ follows from its nullspace. A zero requires $\dot\gamma_{ij}=v_\phi=0$, $v_i^r=\partial_i\theta^r$, $\theta^r=\mathcal{A}s^r$, and $s_T^i=0$. The remaining square then requires $\partial_i s^i=0$ because $\Hc\ne0$. At $k>0$ the longitudinal shift also vanishes, and hence every auxiliary variable and physical velocity vanishes. The auxiliary quadratic form is likewise nonsingular. This gives a direct proof of the all-sector finite-wavelength no-ghost result, including gravitational momentum and lapse constraints, without a short-wavelength approximation.

The gauge-invariant action~\eqref{eq:dual-stueckelberg} also makes the principal propagation transparent. In the original St\"uckelberg variable $\chi^r$, the principal scalar metric is $\operatorname{diag}(1,J^{-1},J^{-1},J^{-1})$, the Maxwell coefficient is $I$, and the Einstein operator is the usual metric wave operator. Derivatives of the coefficients and background field strengths contribute lower-derivative mixing terms. Thus every physical characteristic lies on the metric null cone,
\begin{equation}
 g^{\mu\nu}k_\mu k_\nu=0,\qquad c_{\mathrm{characteristic}}^2=1.
 \label{eq:luminal-characteristics}
\end{equation}
Together with the exact kinetic reduction, this establishes freedom from ghosts and ultraviolet gradient instabilities on Eq.~\eqref{eq:sufficient-no-ghost}. Finite-wavelength evolution is determined by the complete mass and derivative-mixing terms retained in the sector actions. In particular, time-dependent canonical normalization can produce a pump term even when the physical kinetic and principal-gradient matrices are positive.

\subsection{An explicit accelerating homogeneous attractor}
\label{subsec:accelerating-attractor}

The requirement of a positive scalar potential does not prevent the coupled form sectors from changing the asymptotic expansion relative to the scalar-only solution. To establish this point constructively, we now specify a member of the exponential family
\begin{align}
 I&=I_0e^{\alpha\phi/\Mpl},\qquad
 J=J_0e^{\beta\phi/\Mpl},\qquad
 V=V_0e^{-\lambda\phi/\Mpl},\nonumber\\
 \alpha&=\frac{1-\sigma}{3}\lambda,\qquad
 \beta=\frac{2+\sigma}{3}\lambda.
 \label{eq:attractor-family}
\end{align}
where $I_0,J_0,V_0>0$, $\lambda>0$, and $\sigma$ is a dimensionless parameter. In particular, $\alpha+\beta=\lambda$ and $IJV=I_0J_0V_0$. We use this selected family to demonstrate the background dynamics; its Noether-symmetry analysis is deferred to Section~\ref{sec:noether}. No classification of the admissible coupling functions is assumed here.

We use cosmic time, take the expanding branch $\Hc>0$, and introduce the dimensionless variables
\begin{align}
 x&=\frac{\dot\phi}{\sqrt6\Mpl\Hc},\qquad
 y=\frac{\sqrt V}{\sqrt3\Mpl\Hc},\qquad
 \omega=\frac{m}{\Hc\sqrt{IJ}},\nonumber\\
 e&=\frac{\sqrt I\,\dot u}{\sqrt2\Mpl a\Hc},\qquad
 b=\frac{\sqrt J\,\dot v}{\sqrt2\Mpl a^2\Hc}.
 \label{eq:attractor-variables}
\end{align}
Thus $e$ and $b$ are signed amplitudes and $e^2$ and $b^2$ are the corresponding density fractions. The Friedmann constraint, the Hubble slow-roll parameter, and the exponential identity give
\begin{align}
 x^2+y^2+e^2+b^2&=1,\qquad
 \epsilon\equiv-\frac{\dot\Hc}{\Hc^2}=3x^2+2e^2+b^2,\nonumber\\
 \omega&=cy,\qquad
 c\equiv\frac{\sqrt3\Mpl m}{\sqrt{I_0J_0V_0}}.
 \label{eq:attractor-constraints}
\end{align}
Consequently $\omega$ and $y$ are not independent dynamical variables for a specified theory. With a prime denoting $\dd/\dd\ln a$, the background equations become
\begin{subequations}\label{eq:attractor-system}
\begin{align}
 x'&=(\epsilon-3)x+\sqrt{\frac32}\left(\lambda y^2+\alpha e^2+\beta b^2\right),\\
 y'&=\left(\epsilon-\sqrt{\frac32}\lambda x\right)y,\\
 e'&=\left(\epsilon-2-\sqrt{\frac32}\alpha x\right)e+\omega b,\\
 b'&=\left(\epsilon-1-\sqrt{\frac32}\beta x\right)b-\omega e.
\end{align}
\end{subequations}
The opposite signs of the two $BF$ mixing terms follow from the orientation fixed in Eq.~\eqref{eq:bf-contraction}. They also make their contributions to $(e^2+b^2)'$ cancel. The physical phase space for fixed $c$ is three-dimensional; one can eliminate $y=\sqrt{1-x^2-e^2-b^2}$ and $\omega=cy$.

An exact mixed fixed point is obtained for
\begin{equation}
 \sigma=12,\qquad \lambda=\sqrt3,\qquad
 \alpha=-\frac{11}{\sqrt3},\qquad \beta=\frac{14}{\sqrt3},\qquad
 c^2=\frac{102}{83}.
 \label{eq:attractor-parameters}
\end{equation}
For $m>0$, one representative is
\begin{equation}
 (x_*,y_*,e_*,b_*)=
 \left(\frac{1}{\sqrt{18}},\sqrt{\frac{83}{108}},
 \sqrt{\frac{17}{108}},-\frac{1}{\sqrt{54}}\right),
 \qquad \omega_*^2=\frac{17}{18}.
 \label{eq:attractor-fixed-point}
\end{equation}
Changing the signs of both form amplitudes gives the same densities and expansion. The sign of one form amplitude can instead be reversed if the sign of $m$ is reversed. Direct substitution into Eq.~\eqref{eq:attractor-system} gives zero for every right-hand side. The resulting expansion satisfies
\begin{equation}
 \epsilon_* =\frac12,\qquad
 w_{\rm eff}=-1+\frac{2\epsilon_*}{3}=-\frac23,\qquad
 \Omega_{\phi,*}=\frac{89}{108},\quad
 \Omega_{F,*}=\frac{17}{108},\quad
 \Omega_{H,*}=\frac1{54}.
 \label{eq:attractor-observables}
\end{equation}
Here $\Omega_\phi=x^2+y^2$, $\Omega_F=e^2$, and $\Omega_H=b^2$. Both form sectors therefore retain nonzero asymptotic density fractions.

The fixed point corresponds to an exact power-law solution. Let $\tau=t/t_0>0$, with $t_0>0$ and $a_0>0$, and denote $I_c=I(\phi_c)$, $J_c=J(\phi_c)$, and $V_c=V(\phi_c)$. Then
\begin{subequations}\label{eq:attractor-exact-solution}
\begin{align}
 a(t)&=a_0\tau^2, &
 \phi(t)&=\phi_c+\frac{2\Mpl}{\sqrt3}\ln\tau,
 &\Hc(t)&=\frac2t,\\
 I(t)&=I_c\tau^{-22/3},&
 J(t)&=J_c\tau^{28/3},&
 V(t)&=V_c\tau^{-2},\\
 u(t)&=u_c+\frac{2\Mpl a_0}{\sqrt{102I_c}}\tau^{17/3},&
 v(t)&=v_c+\frac{\Mpl a_0^2}{\sqrt{3J_c}}\tau^{-2/3},
\end{align}
\end{subequations}
provided
\begin{equation}
 V_ct_0^2=\frac{83}{9}\Mpl^2,\qquad
 \frac{m^2t_0^2}{I_cJ_c}=\frac{34}{9},\qquad m>0.
 \label{eq:attractor-amplitudes}
\end{equation}
These two relations are compatible precisely with the value of $c^2$ in Eq.~\eqref{eq:attractor-parameters}. The constants $u_c$ and $v_c$ do not enter the field strengths or the stress tensor. In this solution,
\begin{equation}
 \frac{\dot\phi^2}{2}=\frac{2\Mpl^2}{3t^2},\qquad
 V=\frac{83\Mpl^2}{9t^2},\qquad
 \rho_F=\frac{17\Mpl^2}{9t^2},\qquad
 \rho_H=\frac{2\Mpl^2}{9t^2},
 \label{eq:attractor-densities}
\end{equation}
which satisfy the Friedmann, Raychaudhuri, scalar, and both form equations separately.

To assess attraction within the isotropic homogeneous system, linearize the three-dimensional vector field for $(x,e,b)$, keeping $c$ fixed and eliminating $y$ and $\omega$ through Eq.~\eqref{eq:attractor-constraints}. At the fixed point its characteristic polynomial is
\begin{equation}
 \det(s\mathbf1-\mathbf J_*)=
 \frac{(2s+5)(54s^2+135s+1501)}{108}.
 \label{eq:attractor-charpoly}
\end{equation}
The eigenvalues, measured per e-fold of expansion, are
\begin{equation}
 s_1=-\frac52,\qquad
 s_{2,3}=-\frac54\pm\frac{i\sqrt{33999}}{36}.
 \label{eq:attractor-eigenvalues}
\end{equation}
Every eigenvalue has strictly negative real part. The solution is therefore a local asymptotically stable accelerating fixed point of the constrained homogeneous isotropic background dynamics. Since this fixed point is hyperbolic and lies in the interior $y>0$, it also persists, with shifted coordinates and eigenvalues, under sufficiently small changes of the parameters within the family; it is not an isolated nonpersistent balance. If a separately conserved perfect fluid with constant equation of state $w_m$ is added, its density fraction has the linear eigenvalue $2\epsilon_*-3(1+w_m)=-2-3w_m$ about this solution. Dust and radiation perturbations in their background densities thus decay, with eigenvalues $-2$ and $-3$, respectively.

The scalar potential supplies the negative pressure required by Eq.~\eqref{eq:acceleration}, while the coupled forms change the asymptotic expansion. For comparison, the scalar-only exponential solution with the same $\lambda^2=3$ has $w_{\rm eff}=0$~\cite{Copeland1998}. This comparison does not isolate the effect of $BF$ from that of the scalar-dependent form kinetic functions. Keeping the same $I$, $J$, and $V$, including their normalizations and the slopes in Eq.~\eqref{eq:attractor-parameters}, but setting $m=0$ gives $c=\omega=0$. The original form equations, or equivalently Eq.~\eqref{eq:attractor-system} in this limit, admit
\begin{equation}
 x_0=\frac{\sqrt2}{7},\qquad y_0^2=\frac{79}{98},\qquad
 e_0^2=\frac{15}{98},\qquad b_0=0,
 \label{eq:decoupled-accelerating-point}
\end{equation}
with
\begin{equation}
 \epsilon_0=\frac37,\qquad w_{\mathrm{eff},0}=-\frac57,\qquad
 a(t)\propto t^{7/3}.
 \label{eq:decoupled-accelerating-expansion}
\end{equation}
This establishes the existence of an accelerating uncoupled counterpart without invoking the singular $m=0$ limit of the unitary Proca variables. Acceleration is therefore not an exclusive consequence of the topological coupling. In the displayed $m\ne0$ solution, the distinct result is the sustained nonzero density of both form sectors, together with the shifted fixed point $\epsilon_*=1/2$ and $\Omega_{H,*}=1/54$.

\paragraph{Perturbative health of the accelerating background.}
For this solution the dual amplitude and the combinations entering the kinetic matrices are
\begin{align}
 \mathcal{A}&=\frac{2\Mpl a_0}{\sqrt{102I_c}}\tau^{17/3},&
 \frac{I\mathcal{A}^2}{a^2}&=\frac{2\Mpl^2}{51},&
 \frac{gd\mathcal{A}}{h}&=\frac1{18},\\
 \mathcal{U}&=\frac{85\Mpl^2}{9t^2}>0,&
 \mu^2&=\frac{34}{9t^2},&
 \Delta&=h^2g\left(k^2+\frac{76}{81}a^2\mu^2\right)>0.
 \label{eq:attractor-health}
\end{align}
All conditions in Eq.~\eqref{eq:sufficient-no-ghost} hold for every finite $t>0$. The accelerating solution therefore has twelve positive-kinetic modes at every finite $k>0$ and luminal high-frequency characteristics.

Homogeneous anisotropies provide an additional check beyond the isotropic background eigenvalues. Set $T_{ij}=\mathcal{A} z_{ij}$ in the tensor action and seek $(\gamma,z)\propto t^\nu$ at $k=0$. The characteristic equation is
\begin{equation}
 \nu(\nu+5)(27\nu^2+135\nu+76)=0.
\end{equation}
The zero root has $\gamma=2z$ and represents a constant common coordinate deformation with vanishing shear. The physical anisotropy eigenvalues per e-fold are
\begin{equation}
 s=-\frac52,\qquad s=-\frac54\pm\frac{\sqrt{1113}}{36},
 \label{eq:anisotropy-eigenvalues}
\end{equation}
and are all negative. The limiting vector equations for $(X/\mathcal{A},Y/\mathcal{A})$ give the same polynomial. At strictly zero momentum, residual coordinate and orientation freedoms are separated before counting physical homogeneous perturbations.

The antisymmetric helicity-zero mode has the exact decoupled action
\begin{equation}
 S_P^{(2)}=\frac12\int\dd t\,\dd^3k\,aI\left[\dot P^2-\left(\frac{k^2}{a^2}+\mu^2\right)P^2\right].
\end{equation}
With $\dd\eta=\dd t/a$, $\eta=-t_0^2/(a_0t)$, and $v_P=\sqrt I\,P$, its equation becomes
\begin{equation}
 v_P''+\left[k^2+a^2\mu^2-\frac{(\sqrt I)''}{\sqrt I}\right]v_P=0,
 \qquad v_P''+\left(k^2-\frac6{\eta^2}\right)v_P=0.
\end{equation}
Primes in this equation denote conformal-time derivatives. At $k=0$, $P\propto t^{17/3}$ or $t^{2/3}$, while $\mathcal{A}\propto t^{17/3}$. Thus $P/\mathcal{A}$ is constant or decays as $t^{-5}$: the first mode is an infinitesimal constant internal rotation, and the orientation velocity decays. This illustrates how the canonical pump term is compatible with bounded fractional homogeneous deformations. The combined result is a locally attracting accelerating background, with decaying physical homogeneous anisotropy, positive finite-wavelength kinetic matrices, and luminal principal propagation. The exact actions above also specify the finite-wavelength initial-value problem, including its time-dependent mode mixing.

\paragraph{Future boundedness at finite comoving wavelength.}
The exact reduced actions give a further linear-stability result for the accelerating solution. Introduce the dimensionless variables
\begin{equation}
 \ell=\ln(t/t_0),\qquad \kappa=\frac{kt_0^2}{a_0t},\qquad
 \mathbf X=\left(\frac{f}{\Mpl},\frac{T}{\mathcal{A}},\frac{L}{\mathcal{A}}\right)^T,
 \qquad \frac{\dd\kappa}{\dd \ell}=-\kappa.
\end{equation}
After a constant positive normalization is removed, the coupled scalar-block action takes the form
\begin{equation}
 S_{\mathrm{coupled}}^{(2)}=\int\dd \ell\,e^{5\ell}\left[
 \frac12\mathbf X_{,\ell}^T\mathsf K(\kappa)\mathbf X_{,\ell}
 +\mathbf X_{,\ell}^T\mathsf D(\kappa)\mathbf X
 +\frac12\mathbf X^T\mathsf E(\kappa)\mathbf X\right].
\end{equation}
Here $\mathsf K$, $\mathsf D$, and $\mathsf E$ are obtained by substituting the background and the displayed change of variables into the complete Schur-complement action. Thus, with $\mathbf Y=(\mathbf X,\mathbf X_{,\ell})^T$, the exact first-order evolution matrix is
\begin{equation}
 \frac{\dd\mathbf Y}{\dd \ell}=\mathsf M(\kappa)\mathbf Y,\qquad
 \mathsf M(\kappa)=\begin{pmatrix}
 0&\mathbf1\\
 -\mathsf K^{-1}(5\mathsf D-\kappa\mathsf D_{,\kappa}-\mathsf E)&
 -\mathsf K^{-1}(5\mathsf K-\kappa\mathsf K_{,\kappa}+\mathsf D-\mathsf D^T)
 \end{pmatrix}.
\end{equation}
Its entries are rational in $\kappa^2$, with the nonconstant denominator proportional to $729\kappa^2+2584$. In particular, $\mathsf M(\kappa)=\mathsf M_0+O(\kappa^2)$ as $\kappa\to0$, and
\begin{equation}
 \det(\nu\mathbf1-\mathsf M_0)=
 \frac{\nu(\nu+5)(27\nu^2+135\nu+76)
 (27\nu^2+135\nu+3002)}{729}.
\end{equation}
The scalar exponents with respect to $\ell$ are
\begin{equation}
 0,\quad-5,\quad-\frac52\pm\frac{\sqrt{1113}}{18},\quad
 -\frac52\pm\frac{i\sqrt{33999}}{18}.
\end{equation}
The zero root is simple, with representative eigenvector $(1,11\sqrt6/6,11\sqrt3/6,0,0,0)^T$, and gives a constant isotropic adiabatic deformation in spatially flat variables. Every other scalar exponent has negative real part.

The tensor and vector systems, normalized by $(\gamma,T/\mathcal{A})$ and $(X/\mathcal{A},Y/\mathcal{A})$, respectively, have the same rational dependence on $\kappa^2$, no poles at finite nonnegative $\kappa$, and limiting polynomial
\begin{equation}
 \frac{\nu(\nu+5)(27\nu^2+135\nu+76)}{27}.
\end{equation}
Their zero roots are simple and their other roots have negative real part. For the antisymmetric helicity-zero mode, $\Pi_P=P/\mathcal{A}$ obeys $\Pi_{P,\ell\ell}+5\Pi_{P,\ell}+\kappa^2\Pi_P=0$, whose limiting exponents are $0$ and $-5$. Thus each normalized sector has a bounded limiting propagator, with neutral adiabatic, coordinate, or orientation modes and decaying complementary modes.

For every fixed finite comoving $k$ and initial time $t_i>0$, the coefficients are smooth on $t\ge t_i$. Since $\kappa\propto e^{-\ell}$, their late-time correction obeys
\begin{equation}
 \int_{\ell_i}^{\infty}\|\mathsf M(\kappa(\ell))-\mathsf M_0\|\,\dd \ell<\infty.
\end{equation}
Variation of constants, the bounded propagator $e^{\mathsf M_0(\ell-\ell_i)}$, and Gronwall's inequality then give a finite bound on the normalized perturbations and their $\ell$ derivatives. Projection onto the stable and zero-eigenvalue subspaces shows that the stable component decays and the neutral component tends to a finite constant. This establishes future boundedness of all physical linear sectors at each fixed finite comoving wavenumber. The result allows finite transient amplification and does not require every instantaneous canonical frequency to be positive.

\paragraph{Reconstructed constrained fields.}
The eliminated temporal and metric variables can also be reconstructed from the bounded scalar solutions. Define $\widehat S=ktN^z$ and $\widehat A=kt\,\delta C_0^z/\mathcal{A}$. The exact constraint solutions for these variables and the lapse are analytic functions of $\kappa^2$ multiplying $(\mathbf X,\mathbf X_{,\ell})$. The temporal and shift solutions have the same $\kappa=0$ limit for arbitrary normalized scalar state, giving $\widehat A-\widehat S=O(\kappa^2)$. The vector component normal to the spatial slices, relative to the physical background amplitude, therefore satisfies
\begin{equation}
 \frac{\delta C_\perp}{\mathcal{A}/a}
 =\frac{a(\delta C_0^z-\mathcal{A}N^z)}{\mathcal{A}}
 =\frac{\widehat A-\widehat S}{\kappa}=O(\kappa),\qquad\kappa\to0.
\end{equation}
Here $\delta C_\perp=\delta C_0^z-\mathcal{A}N^z$ at linear order, since its background value vanishes. The physical temporal vector thus decays relative to the background even when a coordinate component has a different time dependence. The lapse remains bounded. On the constant adiabatic mode, $\widehat S=O(\kappa^2)$; the other modes decay. More precisely, $\widehat S=O(\tau^{-\gamma_d})$ for any $0<\gamma_d<(45-\sqrt{1113})/18$, so $N^z=\widehat S/(kt)$ is integrable in cosmic time at fixed $k\ne0$. A spatial coordinate transformation removing this shift approaches a finite deformation and gives bounded spatial-metric perturbations. The bounded normalized fields and their derivatives likewise keep the fractional normal electric and spatial Proca perturbations bounded, while spatial-gradient contributions vanish relative to the background as $\kappa\to0$. This reconstruction gives the asymptotic result its physical linear-perturbation interpretation at each fixed finite comoving wavenumber.

\paragraph{Local interaction scales.}
The asymptotic normalization $I\to0$, $J\to\infty$ can be examined in canonically normalized fields. At fixed $\bar\phi$, set $A_c=\sqrt{\bar I}\,A$, $\chi_c=\chi/\sqrt{\bar J}$, and $\varphi=\phi-\bar\phi$. Up to background-derivative terms, the exponential form-sector interactions are
\begin{equation}
 -\frac14e^{\alpha\varphi/\Mpl}F_c^2
 -\frac12e^{-\beta\varphi/\Mpl}(\partial\chi_c-\mu A_c)^2.
\end{equation}
The local interaction coefficients involve $\alpha/\Mpl$ and $\beta/\Mpl$, rather than inverse powers of the background kinetic normalizations. Their parametric interaction scales are $\Mpl/|\alpha|$ and $\Mpl/|\beta|$. On the displayed solution, $\mu=\sqrt{17/18}\,\Hc$, $|\dot I/I|=11\Hc/3$, and $|\dot J/J|=14\Hc/3$, so a local window above the background rates and below these interaction scales opens at sufficiently late times. This is power counting for the stated action; the precise ultraviolet cutoff depends on higher-order operators in its completion.

\section{Noether symmetries and exact solutions}
\label{sec:noether}

We choose a quadratic power law for the effective potential of the pointlike Lagrangian and investigate the zero-boundary Noether symmetries that it admits. The effective potential includes the factor $a^3$ multiplying the canonical scalar potential. Consequently, a quadratic dependence in suitable minisuperspace coordinates does not mean that $V(\phi)$ is a power law in the canonical field. Keeping this distinction explicit, we derive the determining equations, classify their regular branches, and construct an exact accelerating solution with both form field strengths nonzero.

\subsection{Choice of the effective power-law potential}
\label{subsec:noether-potential}

Let $\phi_0$ be a fixed choice of scalar-field origin and introduce
\begin{equation}\begin{gathered}
 \beta_{\mathrm N}\equiv\sqrt{\frac38}\frac1{\Mpl},\qquad
 C_{\mathrm N}\equiv\frac{4\Mpl^2}{3},\qquad
 \theta_{\mathrm N}\equiv\beta_{\mathrm N}(\phi-\phi_0),\\
 Q_1=a^{3/2}\cosh\theta_{\mathrm N},\qquad
 Q_2=a^{3/2}\sinh\theta_{\mathrm N}.
\end{gathered}\label{eq:flat-minisuperspace}
\end{equation}
This is an invertible point transformation on $a>0$, with $Q_1>|Q_2|$ and
\begin{equation}
 a=(Q_1^2-Q_2^2)^{1/3},\qquad
 \phi=\phi_0+\beta_{\mathrm N}^{-1}\operatorname{artanh}(Q_2/Q_1),\qquad
 -3\Mpl^2a\dot a^2+\frac12a^3\dot\phi^2
 =C_{\mathrm N}(-\dot Q_1^2+\dot Q_2^2).
 \label{eq:flat-minisuperspace-inverse}
\end{equation}
The flat two-dimensional kinetic geometry is familiar in canonical scalar minisuperspace~\cite{BenAchour2023Schrodinger}. Related canonical constructions also simplify exponential-potential dynamics through an extended flat-space description~\cite{Paliathanasis2026Hidden}. In these coordinates, Eq.~\eqref{eq:lagrangian} becomes
\begin{equation}
 L=\frac{C_{\mathrm N}}{N}(-\dot Q_1^2+\dot Q_2^2)
 +\frac{3\mathfrak I(Q_1,Q_2)}{2N}\dot u^2
 +\frac{3\mathfrak J(Q_1,Q_2)}{2N}\dot v^2
 -3mv\dot u-N W(Q_1,Q_2),
 \label{eq:noether-flat-L}
\end{equation}
where $\mathfrak I=aI$, $\mathfrak J=J/a$, and $W=a^3V$. For real dimensionless constants $c_1,c_2$ and a positive potential normalization $W_0$, we now make the explicit ansatz
\begin{equation}
 \boxed{W(Q_1,Q_2)=W_0\zeta^2,\qquad \zeta=c_1Q_1+c_2Q_2,\qquad
 W_0>0,\qquad(c_1,c_2)\ne(0,0).}
 \label{eq:chosen-effective-power}
\end{equation}
This is a chosen power-law effective potential, with exponent two; it is not inferred from Noether's theorem. In the original variables it gives
\begin{equation}
 V(\phi)=W_0\left[c_1\cosh\theta_{\mathrm N}+c_2\sinh\theta_{\mathrm N}\right]^2.
 \label{eq:chosen-canonical-potential}
\end{equation}
Thus the canonical scalar potential is hyperbolic-squared, or exponential in a limiting case, rather than a monomial in $\phi$. The positive normalization is also motivated by Eq.~\eqref{eq:acceleration}: with healthy form kinetic signs, a nonpositive potential cannot drive accelerated expansion.

The quadratic choice has a direct structural advantage: it is the simplest nonconstant polynomial in one linear coordinate that is compatible with $W=a^3V(\phi)$. Indeed, under $(Q_1,Q_2)\mapsto(\varsigma Q_1,\varsigma Q_2)$ with a positive scaling factor $\varsigma$ at fixed $\phi$, $a^3\mapsto\varsigma^2a^3$; therefore a single monomial $W_0\zeta^{n_W}$ with exponent $n_W$ must have $n_W=2$ if $V$ is to depend only on $\phi$. For $W_0>0$, its timelike branch gives a positive minimum of the canonical potential, whereas its null branch gives an exponential tail. These provide two distinct mechanisms for potential-supported acceleration, a de Sitter limit and a scaling regime, within a tractable family. Hyperbolic and exponential potentials have also been investigated using Noether methods in scalar cosmology~\cite{Miranda2024Horndeski}; the coupled form degrees of freedom and the velocity-linear $BF$ term must be treated separately here.

Other choices of $V(\phi)$ correspond to other homogeneous functions $W(Q_1,Q_2)$ of degree two and can be studied by the same determining equations. They need not possess the additional autonomous symmetry found below, and their integration may require other point symmetries, further first integrals, approximations, or numerical evolution. The present analysis is restricted to the chosen family; other potential functions require their own symmetry and integrability analysis. Moreover, polynomial simplicity in minisuperspace does not imply a renormalizable polynomial interaction for the canonical scalar.

\subsection{Zero-boundary generator and determining equations}
\label{subsec:noether-equations}

Conserved quantities and adapted cyclic coordinates provide the standard route from Noether symmetries to explicit background solutions~\cite{Giacomini2022Multifield,Laya2023ScalarTensor}. We first classify autonomous configuration-space symmetries that leave $t$ and $N$ unchanged. Their generator and first prolongation are
\begin{align}
 X_{\mathrm N}&=\eta^{a}_{\mathrm N}(a,\phi,u,v)\partial_a+\eta^{\phi}_{\mathrm N}(a,\phi,u,v)\partial_\phi
 +\eta^{u}_{\mathrm N}(a,\phi,u,v)\partial_u+\eta^{v}_{\mathrm N}(a,\phi,u,v)\partial_v,
 \label{eq:noether-vector}\\
 X_{\mathrm N}^{[1]}&=X_{\mathrm N}+\dot\eta^{a}_{\mathrm N}\,\partial_{\dot a}
 +\dot\eta^{\phi}_{\mathrm N}\,\partial_{\dot\phi}
 +\dot\eta^{u}_{\mathrm N}\,\partial_{\dot u}
 +\dot\eta^{v}_{\mathrm N}\,\partial_{\dot v},\qquad
 \dot\eta^{a}_{\mathrm N}=\dot q_{\mathrm N}^{A}\partial_A\eta^{a}_{\mathrm N},
 \label{eq:noether-prolongation}
\end{align}
Here $q_{\mathrm N}^{A}=(a,\phi,u,v)$, $\partial_A\equiv\partial/\partial q_{\mathrm N}^{A}$, and repeated configuration indices are summed. The other dotted coefficients are defined analogously. In the determining equations below, a comma followed by a configuration variable denotes a partial derivative. We impose the strict condition
\begin{equation}
 X_{\mathrm N}^{[1]}L=0,
 \label{eq:strict-noether-condition}
\end{equation}
so the Noether boundary function is identically zero. The representative of the Lagrangian is precisely Eq.~\eqref{eq:lagrangian}, including $-3mv\dot u$; a different representative obtained by integration by parts can change which generators have a zero boundary function. We retain the lapse during the calculation. Since Eq.~\eqref{eq:strict-noether-condition} holds for arbitrary $N$, the resulting symmetries preserve the lapse equation as well as the evolution equations. Cosmic time is imposed only after retaining the Friedmann constraint.

For fixed nonzero $N$, the velocity Hessian is
\begin{equation}
 \mathsf H_{\mathrm N,AB}=\frac{\partial^2L}{\partial\dot q_{\mathrm N}^{A}\partial\dot q_{\mathrm N}^{B}}
 =\frac1N\operatorname{diag}\left(-6\Mpl^2a,a^3,3aI,\frac{3J}{a}\right),
 \qquad
 \det\mathsf H_{\mathrm N}=-\frac{54\Mpl^2a^4IJ}{N^4}.
 \label{eq:noether-hessian}
\end{equation}
It is nondegenerate on every regular patch with $a>0$ and finite positive $I,J$. The negative gravitational entry is the usual minisuperspace signature, not a physical propagating ghost. Including $N$ adds a zero row and column, expressing the lapse constraint. The $BF$ term does not enter this Hessian, although it supplies essential determining equations. In particular, neither the determinant nor the number of Hessian eigenvalues determines the number of symmetry branches.

Equating the coefficients of the ten independent quadratic velocity monomials in Eq.~\eqref{eq:strict-noether-condition} gives
\begin{subequations}\label{eq:noether-quadratic-system}
\begin{align}
 \eta^{a}_{\mathrm N}+2a\eta^{a}_{\mathrm N,a}&=0,&
 3\eta^{a}_{\mathrm N}+2a\eta^{\phi}_{\mathrm N,\phi}&=0,
 \label{eq:noether-base-diagonal}\\
 \eta^{a}_{\mathrm N} I+a\eta^{\phi}_{\mathrm N} I_{,\phi}+2aI\eta^{u}_{\mathrm N,u}&=0,&
 -\eta^{a}_{\mathrm N} J+a\eta^{\phi}_{\mathrm N} J_{,\phi}+2aJ\eta^{v}_{\mathrm N,v}&=0,
 \label{eq:noether-form-diagonal}\\
 a^2\eta^{\phi}_{\mathrm N,a}-6\Mpl^2\eta^{a}_{\mathrm N,\phi}&=0,&
 I\eta^{u}_{\mathrm N,a}-2\Mpl^2\eta^{a}_{\mathrm N,u}&=0,\\
 J\eta^{v}_{\mathrm N,a}-2\Mpl^2a^2\eta^{a}_{\mathrm N,v}&=0,&
 a^2\eta^{\phi}_{\mathrm N,u}+3I\eta^{u}_{\mathrm N,\phi}&=0,\\
 a^4\eta^{\phi}_{\mathrm N,v}+3J\eta^{v}_{\mathrm N,\phi}&=0,&
 a^2I\eta^{u}_{\mathrm N,v}+J\eta^{v}_{\mathrm N,u}&=0.
\end{align}
\end{subequations}
For $m\ne0$, the coefficients linear in velocities give
\begin{equation}
 v\eta^{u}_{\mathrm N,a}=0,\qquad v\eta^{u}_{\mathrm N,\phi}=0,\qquad
 v\eta^{u}_{\mathrm N,v}=0,\qquad \eta^{v}_{\mathrm N}+v\eta^{u}_{\mathrm N,u}=0,
 \label{eq:noether-magnetic-system}
\end{equation}
and the velocity-independent coefficient yields
\begin{equation}
 3\eta^{a}_{\mathrm N} V+a\eta^{\phi}_{\mathrm N} V_{,\phi}=0.
 \label{eq:noether-potential-equation}
\end{equation}
These equations hold as identities on an open configuration-space domain, rather than only along a chosen trajectory. The identities obtained first at $v\ne0$ extend to $v=0$ for smooth generators.

Equation~\eqref{eq:noether-magnetic-system}, followed by the mixed equations in Eq.~\eqref{eq:noether-quadratic-system}, implies
\begin{equation}
 \eta^{u}_{\mathrm N}=c_{\mathrm N} u+d_{\mathrm N},\qquad\eta^{v}_{\mathrm N}=-c_{\mathrm N}v,\qquad
 \eta^{a}_{\mathrm N}=\eta^{a}_{\mathrm N}(a,\phi),\qquad\eta^{\phi}_{\mathrm N}=\eta^{\phi}_{\mathrm N}(a,\phi),
 \label{eq:noether-form-generator}
\end{equation}
where $c_{\mathrm N},d_{\mathrm N}$ are constants. Solving the three gravitational--scalar equations then gives
\begin{equation}
 \eta^{a}_{\mathrm N}=a^{-1/2}f_{\mathrm N}(\phi),\qquad
 \eta^{\phi}_{\mathrm N}=b_{\mathrm N}-4\Mpl^2a^{-3/2}f_{\mathrm N,\phi},\qquad
 f_{\mathrm N,\phi\phi}=\beta_{\mathrm N}^2f_{\mathrm N}.
 \label{eq:noether-base-generator}
\end{equation}
Here $b_{\mathrm N}$ is a constant and $f_{\mathrm N}(\phi)$ is a scalar function. The remaining equations separate into
\begin{align}
 f_{\mathrm N}-4\Mpl^2f_{\mathrm N,\phi}(\ln I)_{,\phi}&=0,&
 b_{\mathrm N}(\ln I)_{,\phi}+2c_{\mathrm N}&=0,\nonumber\\
 -f_{\mathrm N}-4\Mpl^2f_{\mathrm N,\phi}(\ln J)_{,\phi}&=0,&
 b_{\mathrm N}(\ln J)_{,\phi}-2c_{\mathrm N}&=0,\nonumber\\
 3f_{\mathrm N}-4\Mpl^2f_{\mathrm N,\phi}(\ln V)_{,\phi}&=0,&
 b_{\mathrm N}V_{,\phi}&=0.
 \label{eq:noether-separated-system}
\end{align}
For the nonconstant potential chosen in Eq.~\eqref{eq:chosen-canonical-potential}, $b_{\mathrm N}=c_{\mathrm N}=0$. An additional symmetry with $f_{\mathrm N}\ne0$ requires
\begin{equation}
 I\propto V^{1/3},\qquad J\propto V^{-1/3},\qquad
 V\propto (f_{\mathrm N,\phi})^2.
 \label{eq:noether-coupling-result}
\end{equation}
The last relation is a compatibility check on the already chosen potential: its square root is a linear combination of $e^{\beta_{\mathrm N}\phi}$ and $e^{-\beta_{\mathrm N}\phi}$ and hence satisfies the required differential equation. It is not a derivation of the initial potential ansatz. Writing $\mathcal F(\phi)=c_1\cosh\theta_{\mathrm N}+c_2\sinh\theta_{\mathrm N}$, the coupling functions are
\begin{equation}
 I=I_{\mathrm N}|\mathcal F|^{2/3},\qquad J=J_{\mathrm N}|\mathcal F|^{-2/3},\qquad
 I_{\mathrm N},J_{\mathrm N}>0,\qquad IJ=I_{\mathrm N}J_{\mathrm N}.
 \label{eq:noether-couplings}
\end{equation}
Here $I_{\mathrm N}$ and $J_{\mathrm N}$ are the positive coupling normalizations of the present family. The functions are defined on connected regular patches with $\mathcal F\ne0$. In Eq.~\eqref{eq:noether-flat-L}, this result reads $\mathfrak I=I_{\mathrm N}|\zeta|^{2/3}$ and $\mathfrak J=J_{\mathrm N}|\zeta|^{-2/3}$. The two independent generators are therefore
\begin{equation}
 X_{\mathrm N,u}=\partial_u,\qquad X_{\mathrm N,\zeta}=c_2\partial_{Q_1}-c_1\partial_{Q_2}.
 \label{eq:noether-two-generators}
\end{equation}
If the coupling conditions are not satisfied, $X_{\mathrm N,u}$ still remains, but the second generator does not. Equations~\eqref{eq:noether-form-generator}--\eqref{eq:noether-separated-system} establish completeness within the autonomous, lapse-preserving class specified in Eq.~\eqref{eq:noether-vector}. They do not exclude time-dependent or lapse-transforming Noether symmetries.

Related scalar--vector Noether analyses were given by Vakili~\cite{Vakili2014} and Zhang~\cite{Zhang2015}. In particular, the squared linear combination of exponentials in $V$, the relation $I=f^2\propto V^{1/3}$, and the coordinates $Q_1\pm Q_2=a^{3/2}e^{\pm\theta_{\mathrm N}}$ reproduce the structures in Eqs.~(23), (25), and (34) of Ref.~\cite{Zhang2015}, up to constant normalizations and a scalar shift. The present extension includes the two-form condition $J\propto V^{-1/3}$, the $BF$ determining equations, and the coupled-form Bessel solution below. For $m=0$ with the two-form sector removed, $A=\sqrt{3}\,u$ matches the reduced scalar--vector kinetic term; this algebraic correspondence does not identify a single-vector spatial configuration with the isotropic triplet.

Define the canonical momenta by $p_A=\partial L/\partial\dot q_{\mathrm N}^{A}$. The strict Noether integral is $\mathcal Q_{X_{\mathrm N}}=\eta_{\mathrm N}^{A}p_A$, where $\eta_{\mathrm N}^{A}$ denotes the components in Eq.~\eqref{eq:noether-vector}. In particular,
\begin{equation}
 \mathcal Q_u=3\left(\frac{aI\dot u}{N}-mv\right).
 \label{eq:noether-u-charge}
\end{equation}
The other form equation also integrates once,
\begin{equation}
 \frac{J\dot v}{aN}+mu=c_H,
 \label{eq:second-form-first-integral}
\end{equation}
where $c_H$ is the conserved quantity introduced in Eq.~\eqref{eq:dual-background}, now written for arbitrary lapse. The generator $\partial_v$ is not a zero-boundary symmetry of the chosen Lagrangian: its variation is $-3m\dot u$. We use Eq.~\eqref{eq:second-form-first-integral} directly as a field-equation first integral, without assigning it an additional strict symmetry.

\subsection{Three regular branches and cyclic variables}
\label{subsec:noether-branches}

The sign of $D_{\mathrm N}=c_1^2-c_2^2$ divides Eq.~\eqref{eq:chosen-effective-power} into three inequivalent classes. A constant shift of $\phi$, corresponding to a Lorentz transformation in the $(Q_1,Q_2)$ plane, and constant rescalings of the coupling normalizations give the following representatives. In each representative, $\phi_0$ is chosen appropriately in Eq.~\eqref{eq:flat-minisuperspace}; the definitions of $\theta_{\mathrm N}$, $Q_1$, and $Q_2$ are otherwise unchanged. These are branches of the chosen potential and its solved determining equations. The nondegenerate Hessian justifies regular changes of variables on their stated domains; it does not supply this classification.

\paragraph{Branch I: timelike potential coordinate, $D_{\mathrm N}>0$.}
Choose $\zeta=Q_1>0$. Then
\begin{equation}
 V=W_0\cosh^2\theta_{\mathrm N},\qquad
 I=I_{\mathrm N}\cosh^{2/3}\theta_{\mathrm N},\qquad
 J=J_{\mathrm N}\cosh^{-2/3}\theta_{\mathrm N},
 \label{eq:branch-I-functions}
\end{equation}
and the Lagrangian is
\begin{equation}
 L_{\rm I}=\frac{C_{\mathrm N}}{N}(-\dot Q_1^2+\dot Q_2^2)
 +\frac{3I_{\mathrm N}Q_1^{2/3}}{2N}\dot u^2
 +\frac{3J_{\mathrm N}Q_1^{-2/3}}{2N}\dot v^2
 -3mv\dot u-NW_0Q_1^2.
 \label{eq:branch-I-L}
\end{equation}
Both $Q_2$ and $u$ are cyclic, with $p_{Q_2}=2C_{\mathrm N}\dot Q_2/N$. The canonical potential has a regular positive minimum, $V(\phi_0)=W_0$.

\paragraph{Branch II: spacelike potential coordinate, $D_{\mathrm N}<0$.}
Choose $\zeta=Q_2$ and initially take $Q_2>0$; the reflected patch is equivalent. The functions are
\begin{equation}
 V=W_0\sinh^2\theta_{\mathrm N},\qquad
 I=I_{\mathrm N}|\sinh\theta_{\mathrm N}|^{2/3},\qquad
 J=J_{\mathrm N}|\sinh\theta_{\mathrm N}|^{-2/3},
 \label{eq:branch-II-functions}
\end{equation}
with
\begin{equation}
 L_{\rm II}=\frac{C_{\mathrm N}}{N}(-\dot Q_1^2+\dot Q_2^2)
 +\frac{3I_{\mathrm N}Q_2^{2/3}}{2N}\dot u^2
 +\frac{3J_{\mathrm N}Q_2^{-2/3}}{2N}\dot v^2
 -3mv\dot u-NW_0Q_2^2.
 \label{eq:branch-II-L}
\end{equation}
Now $Q_1$ and $u$ are cyclic, with $p_{Q_1}=-2C_{\mathrm N}\dot Q_1/N$. The surface $\theta_{\mathrm N}=0$ is not part of a regular patch: $I$ vanishes and $J$ diverges there. A finite value of the product $IJ$ does not remove this singularity of the individual kinetic coefficients.

\paragraph{Branch III: null potential coordinate, $D_{\mathrm N}=0$.}
Up to $\phi-\phi_0\mapsto-(\phi-\phi_0)$, choose
\begin{equation}
 Q_+=Q_1+Q_2=a^{3/2}e^{\theta_{\mathrm N}},\qquad
 Q_-=Q_1-Q_2=a^{3/2}e^{-\theta_{\mathrm N}},\qquad Q_+,Q_->0.
 \label{eq:null-coordinates}
\end{equation}
The corresponding functions and Lagrangian are
\begin{align}
 V&=W_0e^{2\theta_{\mathrm N}},& I&=I_{\mathrm N}e^{2\theta_{\mathrm N}/3},& J&=J_{\mathrm N}e^{-2\theta_{\mathrm N}/3},
 \label{eq:branch-III-functions}\\
 L_{\rm III}&=-\frac{C_{\mathrm N}}{N}\dot Q_+\dot Q_-
 +\frac{3I_{\mathrm N}Q_+^{2/3}}{2N}\dot u^2
 +\frac{3J_{\mathrm N}Q_+^{-2/3}}{2N}\dot v^2\nonumber\\
 &\quad-3mv\dot u-NW_0Q_+^2.
 \label{eq:branch-III-L}
\end{align}
The cyclic coordinates are $Q_-$ and $u$, and $p_{Q_-}=-C_{\mathrm N}\dot Q_+/N$. A null cyclic direction does not make the kinetic metric degenerate: the $(Q_+,Q_-)$ block has nonzero determinant. Its consequence is instead that the conserved momentum fixes $\dot Q_+/N$ directly, which makes this branch explicitly integrable.

\subsection{Exact solutions and acceleration in the non-null branches}
\label{subsec:nonnull-solutions}

We now use cosmic time, $N=1$, retaining the zero-energy constraint $E_L\equiv\dot q_{\mathrm N}^{A}p_A-L=0$ inherited from lapse variation. In Branch I, the invariant sector $\dot u=\dot v=0$ admits the exact solution
\begin{align}
 Q_1(t)&=A_{\mathrm I,+}e^{\omega_{\mathrm N}t}+A_{\mathrm I,-}e^{-\omega_{\mathrm N}t},&
 Q_2(t)&=Q_{2,0}+b_{\mathrm I}t,&
 \omega_{\mathrm N}^2&=\frac{W_0}{C_{\mathrm N}},\nonumber\\
 a(t)&=[Q_1^2(t)-Q_2^2(t)]^{1/3},&
 \phi(t)&=\phi_0+\frac1{\beta_{\mathrm N}}\operatorname{artanh}\!\frac{Q_2(t)}{Q_1(t)},
 \label{eq:branch-I-exact}\\
 u(t)&=u_c,&v(t)&=v_c,&
 4\omega_{\mathrm N}^2A_{\mathrm I,+}A_{\mathrm I,-}+b_{\mathrm I}^2&=0.
 \nonumber
\end{align}
Here $A_{\mathrm I,+}$, $A_{\mathrm I,-}$, $Q_{2,0}$, $b_{\mathrm I}$, $u_c$, and $v_c$ are real integration constants, and $\omega_{\mathrm N}=\sqrt{W_0/C_{\mathrm N}}>0$. The final equality is the Friedmann constraint, not an optional choice of integration constants. On a future expanding interval with $A_{\mathrm I,+}>0$ and $Q_1>|Q_2|$, this solution approaches
\begin{equation}
 \phi\longrightarrow\phi_0,\qquad
 \Hc\longrightarrow H_0\equiv\frac{2\omega_{\mathrm N}}{3}
 =\sqrt{\frac{W_0}{3\Mpl^2}},\qquad
 a\propto e^{H_0t},\qquad w_{\rm eff}\longrightarrow-1.
 \label{eq:branch-I-desitter}
\end{equation}
These displayed exact trajectories have vanishing form field strengths; the constant potentials $u_c,v_c$ do not constitute an active form energy density.

Within the homogeneous isotropic system, the de Sitter endpoint is also locally stable against excitation of the forms. At the minimum, $I=I_{\mathrm N}$, $J=J_{\mathrm N}$ and $\mu^2=m^2/(I_{\mathrm N}J_{\mathrm N})>0$. The linear scalar equation has $V_{,\phi\phi}(\phi_0)=9H_0^2/4$, so
\begin{equation}
 \delta\phi=(d_{\phi,1}+d_{\phi,2}t)e^{-3H_0t/2}.
 \label{eq:branch-I-scalar-decay}
\end{equation}
With $\bar u\equiv u-c_H/m$, removing the constant form first integrals gives $\ddot{\bar u}+H_0\dot{\bar u}+\mu^2\bar u=0$. Its physical amplitude $\bar u/a$ has exponents
\begin{equation}
 -\frac{3H_0}{2}\pm\sqrt{\frac{H_0^2}{4}-\mu^2},
 \label{eq:branch-I-form-decay}
\end{equation}
with negative real parts; the corresponding orthonormal two-form field-strength components, $H^r_{\hat0\hat i\hat j}=-(m/J_{\mathrm N})(\bar u/a)\epsilon^r{}_{ij}$, and energy density $\rho_H$ decay as well. The constants $d_{\phi,1}$ and $d_{\phi,2}$ specify the scalar perturbation amplitude. The double-root cases are understood with the usual polynomial prefactor. Thus small homogeneous form perturbations do not prevent the late de Sitter limit. Together with $V>0$ and positive $I,J$, this identifies Branch I as a regular accelerating candidate in the kinetic and background sense specified in Section~\ref{sec:perturbations}. It is not a fit to late-time cosmological data.

For comparison, an exact field-strength-free family in Branch II is
\begin{align}
 Q_1(t)&=Q_{1,0}+b_{\mathrm{II}}t,& Q_2(t)&=A_{\mathrm{II}}\cos(\omega_{\mathrm N}t)+B_{\mathrm{II}}\sin(\omega_{\mathrm N}t),
 \nonumber\\
 b_{\mathrm{II}}^2&=\omega_{\mathrm N}^2(A_{\mathrm{II}}^2+B_{\mathrm{II}}^2),&u(t)&=u_c,\qquad v(t)=v_c,
 \label{eq:branch-II-exact}
\end{align}
Here $Q_{1,0}$, $b_{\mathrm{II}}$, $A_{\mathrm{II}}$, and $B_{\mathrm{II}}$ are real integration constants; $a$ and $\phi$ are recovered from Eq.~\eqref{eq:flat-minisuperspace-inverse}. The solution is restricted to $Q_1>|Q_2|$ and $Q_2\ne0$. Its nontrivial oscillating $Q_2$ reaches zero at finite times, so each solution of the full regular theory is confined to a connected interval avoiding those zeros. More generally, even with both form sectors active, the cyclic coordinate in Eq.~\eqref{eq:branch-II-L} enforces $Q_1=Q_{1,0}+b_{\mathrm{II}}t$. Every physical trajectory therefore satisfies
\begin{equation}
 0<a^3=Q_1^2-Q_2^2\leq(Q_{1,0}+b_{\mathrm{II}}t)^2.
 \label{eq:branch-II-growth-bound}
\end{equation}
A future-complete expanding solution cannot remain accelerating at all sufficiently late times: $\ddot a>0$ and $\dot a>0$ from some time onward would imply at least linear growth, in conflict with $a=O(t^{2/3})$. Branch II can have transient acceleration, but it cannot provide sustained late-time acceleration on a regular future-complete expanding trajectory.

The cyclic variables do not by themselves prove complete integrability of the generic active-form dynamics in Branch I or II. To display what remains, put $\zeta=Q_1,\zeta_{\mathrm{cyc}}=Q_2,\epsilon_{\mathrm N}=-1$ in Branch I and $\zeta=Q_2,\zeta_{\mathrm{cyc}}=Q_1,\epsilon_{\mathrm N}=+1$ in Branch II, on a patch with $\zeta>0$. Eliminating $u$ at fixed $p_u=3\mathcal P_u=\mathcal Q_u$ and $\zeta_{\mathrm{cyc}}$ at fixed $p_{\zeta_{\mathrm{cyc}}}=\mathcal Q_{\mathrm{cyc}}$ gives the Routhian
\begin{equation}
 L_{\rm red}=\epsilon_{\mathrm N} C_{\mathrm N}\dot \zeta^2
 +\frac{3J_{\mathrm N}}{2}\zeta^{-2/3}\dot v^2
 -\frac{3(\mathcal P_u+mv)^2}{2I_{\mathrm N}\zeta^{2/3}}-W_0\zeta^2
 +\frac{\mathcal Q_{\mathrm{cyc}}^2}{4\epsilon_{\mathrm N} C_{\mathrm N}}.
 \label{eq:nonnull-routhian}
\end{equation}
The last term is constant. The coupled $(\zeta,v)$ evolution and the original zero-energy constraint must still be solved. Equations~\eqref{eq:branch-I-exact} and~\eqref{eq:branch-II-exact} are exact invariant-sector solutions, rather than a claimed general solution of this remaining system.

\subsection{An exact accelerating solution with active BF-coupled forms}
\label{subsec:null-exact}

Branch III admits a complete explicit integration of the expanding homogeneous system. Its cyclic momentum fixes
\begin{equation}
 Q_+=\upsilon_+\mathcal T,\qquad\mathcal T\equiv t-t_{\mathrm B}>0,\qquad
 \upsilon_+>0,\qquad \mathcal Q_-\equiv p_{Q_-}=-C_{\mathrm N}\upsilon_+.
 \label{eq:null-R-linear}
\end{equation}
Here $t_{\mathrm B}$ is a time-origin constant and $\upsilon_+$ is the constant positive velocity of $Q_+$. The two form first integrals may be written as
\begin{equation}
 I_{\mathrm N}Q_+^{2/3}\dot u=m(v-v_c),\qquad
 J_{\mathrm N}Q_+^{-2/3}\dot v=-m(u-u_c),
 \label{eq:null-form-first-integrals}
\end{equation}
where $u_c=c_H/m$ and $v_c=-\mathcal Q_u/(3m)$ are constants. Hence, with $\bar u=u-u_c$ and $\mu=|m|/\sqrt{I_{\mathrm N}J_{\mathrm N}}$,
\begin{equation}
 \ddot{\bar u}+\frac{2}{3\mathcal T}\dot{\bar u}+\mu^2\bar u=0.
 \label{eq:null-bessel-equation}
\end{equation}
Define
\begin{equation}
 \mathcal Z_{\varrho}(\mathcal T)=A_{\mathrm B}\,\mathrm J_{\varrho}(\mu\mathcal T)+B_{\mathrm B}\,\mathrm Y_{\varrho}(\mu\mathcal T),
 \label{eq:null-bessel-definition}
\end{equation}
where $\varrho$ denotes the order, $\mathrm J_{\varrho}$ and $\mathrm Y_{\varrho}$ are Bessel functions of the first and second kind, and the same real constants $A_{\mathrm B},B_{\mathrm B}$ are used at every order. The exact form solution is
\begin{equation}
 \boxed{\begin{aligned}
 u(t)&=u_c+\mathcal T^{1/6}\mathcal Z_{1/6}(\mathcal T),\\
 v(t)&=v_c+\frac{I_{\mathrm N}\mu\upsilon_+^{2/3}}{m}\,
 \mathcal T^{5/6}\mathcal Z_{-5/6}(\mathcal T).
 \end{aligned}}
 \label{eq:null-form-solution}
\end{equation}
It follows directly from $\dd[\mathcal T^{\varrho}\mathcal Z_{\varrho}(\mathcal T)]/\dd\mathcal T=\mu\mathcal T^{\varrho}\mathcal Z_{\varrho-1}(\mathcal T)$, with the argument convention of Eq.~\eqref{eq:null-bessel-definition}. Both field strengths are generically nonzero when $(A_{\mathrm B},B_{\mathrm B})\ne(0,0)$, and their evolution depends explicitly on $m$.

In expressions below, $\mathcal Z_{\varrho}$ without an argument means $\mathcal Z_{\varrho}(\mathcal T)$. The positive form kinetic contribution to the minisuperspace energy is
\begin{align}
 K_{\mathrm{forms}}&\equiv\frac{3I_{\mathrm N}Q_+^{2/3}}{2}\dot u^2
 +\frac{3J_{\mathrm N}Q_+^{-2/3}}{2}\dot v^2\nonumber\\
 &=\frac{3I_{\mathrm N}\mu^2\upsilon_+^{2/3}}{2}\,
 \mathcal T\left[\mathcal Z_{-5/6}^2+\mathcal Z_{1/6}^2\right].
 \label{eq:null-form-energy}
\end{align}
The lapse equation becomes
\begin{equation}
 -C_{\mathrm N}\upsilon_+\dot Q_-+K_{\mathrm{forms}}+W_0\upsilon_+^2\mathcal T^2=0.
 \label{eq:null-constraint}
\end{equation}
A useful explicit primitive is
\begin{equation}
 \mathcal B_{\mathrm N}(\mathcal T)=I_{\mathrm N}\upsilon_+^{2/3}\left\{
 \frac{3\mu^2\mathcal T^2}{2}\left[\mathcal Z_{-5/6}^2+\mathcal Z_{1/6}^2\right]
 +\mu\mathcal T \mathcal Z_{-5/6}\mathcal Z_{1/6}\right\},\qquad
 \frac{\dd\mathcal B_{\mathrm N}}{\dd\mathcal T}=K_{\mathrm{forms}}.
 \label{eq:null-energy-primitive}
\end{equation}
Thus no unevaluated quadrature is needed:
\begin{equation}
 Q_-(\mathcal T)=Q_{-,c}+\frac1{C_{\mathrm N}\upsilon_+}
 \left[\frac{W_0\upsilon_+^2\mathcal T^3}{3}+\mathcal B_{\mathrm N}(\mathcal T)\right],
 \label{eq:null-S-solution}
\end{equation}
where $Q_{-,c}$ is a real integration constant. The cosmological variables are
\begin{equation}
 \boxed{a(t)=[\upsilon_+\mathcal T Q_-(\mathcal T)]^{1/3},\qquad
 \phi(t)=\phi_0+\frac1{2\beta_{\mathrm N}}\ln\frac{\upsilon_+\mathcal T}{Q_-(\mathcal T)}.}
 \label{eq:null-cosmological-solution}
\end{equation}
Equations~\eqref{eq:null-R-linear}, \eqref{eq:null-form-solution}, \eqref{eq:null-S-solution}, and~\eqref{eq:null-cosmological-solution} give the full constrained expanding family in this branch. The constants are $\upsilon_+,Q_{-,c},A_{\mathrm B},B_{\mathrm B},u_c,v_c,t_{\mathrm B}$, with the restrictions stated above and below; $\phi_0$ labels the chosen coupling functions. The remaining $Q_+$ equation follows by differentiating the constraint and using the form equations and $\ddot Q_+=0$, since $\dot Q_+=\upsilon_+\ne0$.

Locally, $u_c$ and $v_c$ specify gauge-dependent origins of the form potentials: constant shifts of $u$ and $v$ induce the same shifts of $u_c$ and $v_c$, leaving $u-u_c$, $v-v_c$, and both field strengths unchanged. The global admissibility of these shifts depends on the spatial topology and boundary conditions, so the listed integration constants should not be counted as independent physical parameters without fixing this information.

Choose any $\mathcal T_i>0$ for which $Q_-(\mathcal T_i)>0$. Equation~\eqref{eq:null-constraint} then gives $\dot Q_->0$, so $Q_+,Q_-$ remain positive and the solution exists on the full future interval $\mathcal T\geq\mathcal T_i$. Moreover,
\begin{equation}
 \Hc=\frac13\left(\frac1{\mathcal T}+\frac{\dot Q_-}{Q_-}\right)>0.
 \label{eq:null-Hubble}
\end{equation}
All couplings are finite and positive at every finite future time. The case $\upsilon_+=0$ is excluded by the constraint with $W_0>0$, because it would require $K_{\mathrm{forms}}+W_0Q_+^2=0$.

The large-argument Bessel expansion yields $u-u_c=O(\mathcal T^{-1/3})$, $v-v_c=O(\mathcal T^{1/3})$, $K_{\mathrm{forms}}=O(1)$, and $\mathcal B_{\mathrm N}=O(\mathcal T)$. Consequently,
\begin{align}
 Q_-(\mathcal T)&=\frac{W_0\upsilon_+}{3C_{\mathrm N}}\mathcal T^3+O(\mathcal T),\nonumber\\
 a(t)&=\left(\frac{W_0\upsilon_+^2}{3C_{\mathrm N}}\right)^{1/3}
 \mathcal T^{4/3}\left[1+O(\mathcal T^{-2})\right],\nonumber\\
 \phi(t)&=\phi_0+\frac1{2\beta_{\mathrm N}}\ln\frac{3C_{\mathrm N}}{W_0\mathcal T^2}
 +O(\mathcal T^{-2}),
 \label{eq:null-asymptotic-solution}\\
 \Hc&\sim\frac4{3\mathcal T},\qquad
 w_{\rm eff}\longrightarrow-\frac12,\qquad
 \frac{\ddot a}{a}\sim\frac4{9\mathcal T^2}>0.
 \nonumber
\end{align}
Although the coordinate amplitude $v$ can grow, its physical energy density decays:
\begin{equation}
 \rho_F+\rho_H=\frac{K_{\mathrm{forms}}}{a^3}=O(\mathcal T^{-4}),\qquad
 \Omega_F+\Omega_H=O(\mathcal T^{-2}),\qquad
 I=O(\mathcal T^{-2/3}),\quad J=O(\mathcal T^{2/3}).
 \label{eq:null-form-dilution}
\end{equation}
The BF coupling controls the oscillations and the finite-time backreaction, while the leading accelerating exponent coincides with that of the scalar exponential solution with $\lambda^2=3/2$~\cite{Copeland1998}. Thus the acceleration is supplied by the chosen scalar potential and is not attributed to a direct negative-pressure contribution from the topological term.

For this expanding family, $V>0$, $I,J>0$, and $\mathcal U=V+\rho_H>0$. It therefore satisfies the sufficient no-ghost conditions of Eq.~\eqref{eq:sufficient-no-ghost}, and its principal characteristics are luminal. These statements establish an exact, regular, kinetically healthy late-time accelerating family with nonzero BF-coupled fields. They do not by themselves establish observational agreement or future boundedness of every finite-wavelength perturbation. The detailed eigenvalues and boundedness proof for the different fixed point in Section~\ref{subsec:accelerating-attractor} cannot be transferred to this solution without evaluating its own perturbation equations.

\subsection{Relation to the earlier exponential model}
\label{subsec:noether-exponential-relation}

The quadratic effective-potential branches above have $IJ=I_{\mathrm N}J_{\mathrm N}$. The exponential family in Eq.~\eqref{eq:attractor-family} instead has $IJV=I_0J_0V_0$ and a nonconstant $IJ$. It is therefore a distinct family, rather than another representative of Branch III. Its symmetry involves a transformation of time, which lies outside the autonomous class of Eq.~\eqref{eq:noether-vector}.

For a point generator $\widehat X_{\mathrm N}=\tau_{\mathrm N}(t)\partial_t+\widehat\eta_{\mathrm N}^{A}(q_{\mathrm N})\partial_A$, where $\tau_{\mathrm N}$ is its time component and $\widehat\eta_{\mathrm N}^{A}$ are its configuration components, the zero-boundary condition is
\begin{equation}
 \widehat X_{\mathrm N}^{[1]}L+L\dot\tau_{\mathrm N}=0,\qquad
 \widehat X_{\mathrm N}^{[1]}=\widehat X_{\mathrm N}+
 \left(\dot{\widehat\eta}_{\mathrm N}^{A}-\dot q_{\mathrm N}^{A}\dot\tau_{\mathrm N}\right)\partial_{\dot q_{\mathrm N}^{A}}.
 \label{eq:time-noether-condition}
\end{equation}
In cosmic time, direct substitution of Eq.~\eqref{eq:attractor-family} gives
\begin{equation}
 \widehat X_{\mathrm N}=t\partial_t+\frac a3\partial_a
 +\frac{2\Mpl}{\lambda}\partial_\phi
 +\frac\sigma3\left(u\partial_u-v\partial_v\right).
 \label{eq:exponential-time-generator}
\end{equation}
The associated integral is
\begin{equation}
 \widehat{\mathcal Q}_{\mathrm N}=\frac a3p_a+\frac{2\Mpl}{\lambda}p_\phi
 +\frac\sigma3(up_u-vp_v)-tE_L,
 \qquad E_L=\dot q_{\mathrm N}^{A}p_A-L=0.
 \label{eq:exponential-noether-charge}
\end{equation}
Equivalently, before fixing the lapse, its configuration-space part together with $N\partial_N$ is a strict autonomous symmetry of Eq.~\eqref{eq:lagrangian}. This verifies the symmetry of the previously selected exponential model without identifying its mixed fixed point with the new constant-$IJ$ solutions.

Within the chosen quadratic effective-potential family, Branch I supports a regular de Sitter limit, Branch II is excluded from sustained future acceleration by its cyclic-coordinate growth bound, and Branch III supplies an explicit accelerating solution with both form sectors active. The distinction between these branches follows from the solved determining equations and the causal type of the linear potential coordinate. The Hessian establishes regularity, and the Friedmann constraint and perturbative conditions provide independent tests of the resulting cosmologies.

\section{Cosmological evolution of the exact solutions}
\label{sec:numerics}

\begin{figure}[t]
 \centering
 \includegraphics[width=\textwidth]{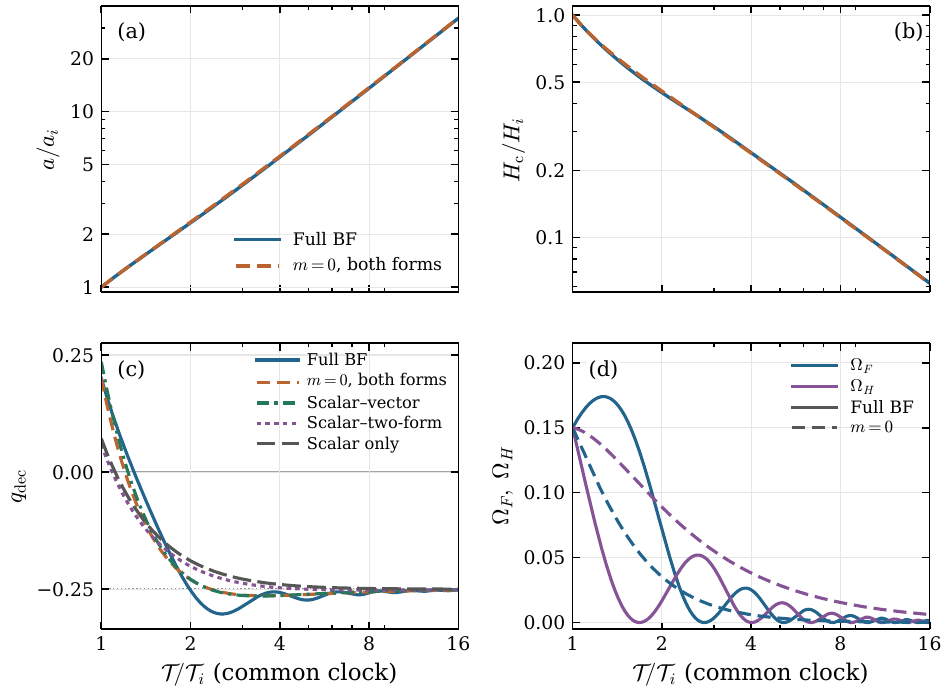}
 \caption{Branch III and its consistent limits. Panels (a) and (b) compare $a/a_i$ and $\Hc/H_i$ for the full model ($m/H_i=1$, solid) and the matched $m=0$ model (dashed). Panel (c) shows $q_{\rm dec}$ for all five cases; the dotted horizontal line marks $-1/4$. Panel (d) separates $\Omega_F$ and $\Omega_H$ for the full and decoupled models. We set $I_{\mathrm N}=J_{\mathrm N}=1$, $W_0/(\Mpl^2H_i^2)=27/20$, $a_i=1$, $\phi_i=\phi_0=u_i=v_i=0$, $\dot\phi_i/(\Mpl H_i)=-\sqrt{3/2}$, and each retained form velocity equal to $\sqrt{3/10}\,\Mpl H_i$. The common clock is $\mathcal T/\mathcal T_i=1+3H_i(t-t_i)/4$, with $\mathcal T_i=4/(3H_i)$ fixed by the full solution, rather than reset for each truncation. All horizontal axes and the vertical axes in (a) and (b) are logarithmic.}
 \label{fig:null-evolution}
\end{figure}

We follow the cosmic-time evolution of the exact Branch III solution and its consistent limits. Figure~\ref{fig:null-evolution} shows the expansion history, the deceleration parameter, and the separate form energy fractions. We define
\begin{equation}
 q_{\rm dec}\equiv-1-\frac{\dot\Hc}{\Hc^2},\qquad
 \Omega_\phi\equiv\frac{\dot\phi^2/2+V}{3\Mpl^2\Hc^2},\qquad
 \Omega_A\equiv\frac{\rho_A}{3\Mpl^2\Hc^2}\quad(A=F,H),
 \label{eq:evolution-observables}
\end{equation}
so that $\Omega_\phi+\Omega_F+\Omega_H=1$ and acceleration corresponds to $q_{\rm dec}<0$. Here $H_i$ denotes the initial Hubble rate of the full model and is the common reference scale for every case. With the initial data in the caption, the full and decoupled models have
\begin{equation}
 \frac{\dot\phi_i^2}{6\Mpl^2H_i^2}=\frac14,\qquad
 \frac{V(\phi_i)}{3\Mpl^2H_i^2}=\frac9{20},\qquad
 \Omega_{F,i}=\Omega_{H,i}=\frac3{20},\qquad q_{{\rm dec},i}=\frac15.
 \label{eq:evolution-initial-fractions}
\end{equation}
\subsection{Comparison with consistent limiting cases}
\label{subsec:consistent-limit-comparison}

We compare the exact Branch III background with four consistent limits: the decoupled model with $m=0$ and both forms retained; the scalar--vector model with $m=0$ and $H^r_{\mu\nu\rho}=0$; the scalar--two-form model with $m=0$ and $F^r_{\mu\nu}=0$; and the scalar-only sector with both field strengths zero. The last restriction is consistent for any $m$, whereas removing just one active form at $m\ne0$ violates the equation of the omitted sector. All cases use the same Branch III potential and the same surviving kinetic functions in Eq.~\eqref{eq:branch-III-functions}.

For each truncation, the surviving fields and their physical initial velocities are unchanged; the removed field strength is set to zero. The expanding initial Hubble rate must then be recomputed from Eq.~\eqref{eq:friedmann}. Thus $\Hc(t_i)/H_i=\sqrt{17/20}$ for either single-form case and $\sqrt{7/10}$ for the scalar-only case. Their initial total energies differ, so these comparisons illustrate the role of each sector; only the matched full and decoupled pair isolates the effect of the BF interaction.

The full solution follows Eqs.~\eqref{eq:null-form-solution}--\eqref{eq:null-cosmological-solution}. At $m=0$, the conserved quantities $aI\dot u$ and $J\dot v/a$ give the form velocities directly, and Eq.~\eqref{eq:null-constraint} determines $Q_-$. Setting either or both conserved form momenta to zero gives the three truncations, without a singular substitution in the Bessel constants. In each case $Q_+(t_i)=Q_-(t_i)=1$ and $\upsilon_+=3\Hc(t_i)/2+\beta_{\mathrm N}\dot\phi_i>0$, so the constraint preserves $Q_\pm>0$ and $I,J>0$. All curves are checked independently against the original cosmic-time background equations. The reconstructed Friedmann residual and conserved-quantity drifts are below $10^{-12}$; direct integration agrees to better than $10^{-8}$ in the dimensionless units of the figure.

The full and decoupled models begin accelerating at $H_i(t-t_i)\simeq0.386$ and $0.269$, respectively. The corresponding values are $0.307$ for scalar--vector, $0.117$ for scalar--two-form, and $0.151$ for scalar-only. The BF coupling produces oscillatory transfer between the two form sectors and damped oscillations in $q_{\rm dec}$. Nevertheless, over $0\leq H_i(t-t_i)\leq20$, the maximum full-versus-decoupled differences are only $1.3\%$ in $a$ and $2.0\%$ in $\Hc$. At the endpoint, the full model has $\Omega_F+\Omega_H\simeq1.57\times10^{-3}$ and $q_{\rm dec}\simeq-0.2515$. These modest expansion corrections characterize this benchmark, not a general bound on BF effects.

The unequal pressures, $p_F=\rho_F/3$ and $p_H=-\rho_H/3$, explain why energy transfer between the forms affects acceleration without a direct BF stress tensor. In this benchmark, the early transfer toward the one-form sector raises the form pressure and contributes to the later onset of acceleration. The scalar-dependent couplings also exchange energy with the scalar, so the full solution includes both processes before the form fractions decline and the scalar potential controls the late expansion.

All five trajectories approach the scalar-potential-driven limit $a\propto t^{4/3}$, $q_{\rm dec}\to-1/4$, and $w_{\rm eff}\to-1/2$. For the coupled solution, Eqs.~\eqref{eq:null-asymptotic-solution} and~\eqref{eq:null-form-dilution} give $\Omega_F+\Omega_H=O(\mathcal T^{-2})$. In the decoupled model, Eq.~\eqref{eq:decoupled-first-integrals} instead gives $\rho_F=O(t^{-14/3})$ and $\rho_H=O(t^{-10/3})$, hence $\Omega_H=O(t^{-4/3})$. Its two-form fraction thus decays more slowly than the combined coupled-form fraction. The scalar-only comparison confirms that this accelerating exponent already follows from the same Branch III exponential potential; the BF interaction modifies the finite-time evolution and form-energy distribution. The field-strength-free Branch I solution instead approaches the de Sitter limit $\Hc\to H_0$ and $q_{\rm dec}\to-1$ of Eq.~\eqref{eq:branch-I-desitter}. These trajectories illustrate the dynamics and do not constitute an observational fit. Recent tests of exponential and related quintessence potentials show why background acceleration alone does not establish a preference over $\Lambda$CDM~\cite{Ramadan:2024kmn,Bayat:2025xfr}.

\section{Conclusions}
\label{sec:conclusions}
\clubpenalty=10000
\widowpenalty=10000

We studied an isotropic scalar--vector--two-form cosmology with a constant topological $BF$ coupling. The lapse-preserving reduction gives the Friedmann constraint together with the scalar and form equations, and the local dual description fixes the number of physical degrees of freedom at twelve. Positive kinetic functions, together with the sufficient background conditions derived here, give positive kinetic matrices at finite wavelength and luminal principal propagation. For the separate exponential family of Subsection~\ref{subsec:accelerating-attractor}, we found an exact mixed accelerating solution with $a\propto t^2$. It is a local attractor of the constrained isotropic system, and its normalized physical linear perturbations remain bounded at each fixed finite comoving wavenumber.

For the Noether analysis, we chose a quadratic effective potential in the flat gravitational--scalar minisuperspace. This choice precedes the symmetry calculation; the corresponding canonical potential is hyperbolic-squared or exponential. The zero-boundary determining equations fix the compatible kinetic functions and separate the solutions into three branches. The velocity Hessian checks regularity, while the causal character of the potential coordinate distinguishes the branches. Branch I admits a regular de Sitter limit, whereas the cyclic-coordinate bound excludes sustained acceleration on regular future-complete expanding trajectories in Branch II. In Branch III, the cyclic variables yield explicit scale-factor and field solutions with both $BF$-coupled form sectors active. Its regular expanding family approaches $a\propto t^{4/3}$ and $w_{\rm eff}\to-1/2$, while the total form fraction decays as $\mathcal T^{-2}$.

The consistent-limit comparison clarifies the source of this acceleration. The scalar potential drives the late expansion, and the topological interaction changes the exchange and finite-time backreaction of the form fields. Its effect on the scale factor and Hubble rate is modest for the matched benchmark considered here. Removing either form sector requires $m=0$ unless both field strengths vanish; the reduced models must also satisfy their own Friedmann constraints. These distinctions are necessary when comparing their early evolution.

The exact solutions provide controlled backgrounds for further tests. The finite-wavelength boundedness result for the mixed exponential fixed point does not establish the same result for every Noether branch. A separate perturbation analysis of Branch III, together with matter and radiation and a comparison with cosmological data, is needed to assess its role as a late-time dark-energy model.

\begingroup
\raggedright
\bibliographystyle{unsrt}
\bibliography{references}
\endgroup
\end{document}